\documentclass[11pt]{article}
\usepackage{amsmath,amssymb}
\usepackage{framed}

\usepackage{graphicx}

\usepackage{amsmath,latexsym}

\begin{document}

\setcounter{page}{1}

\pagestyle{plain}

\begin{center}
	\Large{\bf Viable Anisotropic Inflation and Reheating in the Tachyon Model}\\
	\small \vspace{1cm} {\bf Narges Rashidi $^{}$\footnote{n.rashidi@umz.ac.ir }}  \\
	\vspace{0.5cm} Department of Theoretical Physics, Faculty of
	Science,
	University of Mazandaran,\\
	P. O. Box 47416-95447, Babolsar, IRAN\\
\end{center}

\begin{abstract}
	We study the intermediate tachyon inflation in an anisotropic
	background. By using the Friedmann equations obtained in the
	anisotropic geometry, we obtain the slow-roll parameters in the
	tachyon model. The presence of the anisotropic effects in the
	slow-roll parameters changes the perturbation parameters in our
	setup which may change its observational viability. To check this,
	we perform a numerical analysis and test the results with Planck2018
	TT, TE, EE +lowE+lensing+BK14(18)+BAO data. We show that the
	intermediate anisotropic inflation in some ranges of the anisotropic
	and intermediate parameters is observationally viable. We also show
	that the equilateral amplitude of the non-gaussianity in our model
	is of the order of $10^{-2}-10^{-1}$. By studying the reheating
	process in our setup, we find that it is possible to have
	instantaneous reheating in this model. We also find that the
	temperature during the reheating in our setup is consistent with
	Big-Bang nucleosynthesis.
\end{abstract}
\newpage

\section{Introduction}
Inflation is an interesting paradigm that has solved some problems
of the standard theory of cosmology in a logical way. More
importantly, this paradigm which has been confirmed by several
cosmic microwave background radiation probes, gives a fine
explanation for the origin of the large-scale structure in the
universe. In this regard, different kinds of the inflation model
have been proposed~\cite{Gut81,Lin82,Alb82,Lin90,Lid00a,Lid97}.
Although the simple single field inflation model leads to the scale
invariant and gaussian distributed amplitude of the
non-gaussianity~\cite{Mal03}, there are some models that predict the
non-gaussian features of the perturbations. Tachyon
Inflation~\cite{Bil18,Ras21,Ras22}, DBI
inflation~\cite{Cai11,Raso19}, Gauss-Bonnet
inflation~\cite{Kan15,Noj19}, non-minimal~\cite{Far96,Far00}, and
non-minimal derivative inflation~\cite{Yan15,Noz18} are some of the
models having the capability to give non-gaussian primordial
perturbations. Among these models, in this work, we are interested
in the tachyon inflation. This field gives the possibility to have a
universe evolving from the accelerating phase of the expansion to
the era dominated by the non-relativistic
fluid~\cite{Sen99,Sen02a,Sen02b,Gib02}. Working on the models with
the tachyon field leads to interesting cosmological implications.

Most inflationary models are based on the cosmological principle. It
means the background in these models is homogeneous and isotropic
described by the FRW metric. However, it is believed that at very
primordial instants, the physical description of our universe may
not be just a simple isotropic model. On the other hand, the recent
observations release some anomalies of the cosmic microwave
background radiation~\cite{Kom11,Hin12,Pl18a,Pl18b,Pl18c}. Although
some researchers have proposed instrumental origin for these
anomalies rather than the
cosmological~\cite{Han09,Han10,Ben12,Kim13}, there are some
cosmologists who considered a homogeneous but anisotropic background
in the early universe~\cite{Ell69,Gum07,Pit08}. The cosmic no-hair
conjecture predicts, that even if the universe started from an
anisotropic state, it eventually approaches a homogeneous and
isotropic state~\cite{Gib77,Haw82}. Also, it is shown that this
conjecture can be locally valid~\cite{Sta83,Mul90,Bar84,Ste87}.
Though, in Ref.~\cite{Col19} the authors have presented the
possibility that the late-time universe might be anisotropic. Some
authors have also presented counterexamples against the validity of
the no-hair conjecture~\cite{Bar06a,Bar06b,Mid10,Mul18}. Yet, it has
been shown that, during the inflation era, some of these
counterexamples are unstable~\cite{Kao09,Kao11a,Kao11b}. There is
also a counterexample to the conjecture found in the model based on
supergravity~\cite{Wat09,Kan10}.

In this regard, some interesting works on anisotropic inflation have
been done. For example, the authors of Ref.~\cite{Bar10}, by
considering additional quadratic Ricci curvature terms in the
Einstein-Hilbert have studied some cosmological solutions. In
Ref.~\cite{Dul10}, by adopting a special coupling between the gauge
field and the inflaton, the primordial gravitational wave and the
power spectral in the anisotropic inflation have been studied. The
authors of Ref.~\cite{Kan10} have shown that in the case of
exponential potential for both inflaton and gauge field, there exist
exact anisotropic power-law inflationary solutions. In
Ref.~\cite{Lah16} it has been shown that, by considering an
exponential function for inflaton potential, the Gauss-Bonnet
coupling term, and the gauge coupling term, we get an anisotropic
power-law inflation. The anisotropic constant-roll inflation has
been studied in Ref.~\cite{Ito18}. The authors of~\cite{Do21a}, have
considered the non-canonical anisotropic inflation and studied the
scalar and tensor perturbations in that setup. In Ref.~\cite{Do21b},
the authors have considered a model where there are non-minimal
couplings between two scalar fields and two vector fields. They have
solved the anisotropic power-law solution in their setup. The
authors of Ref.~\cite{Che21} have considered hyperbolic inflation
along a gauge field and have studied anisotropic inflation in their
setup. Another interesting work on anisotropic inflation has been
done in Ref.~\cite{Noj22}. In that paper, the authors have
considered a general form of the metric, based on the Bianchi IX
cosmology, and have studied the anisotropic inflation in the $F(R)$
gravity.

In this paper, following Ref.~\cite{Noj22}, we adopt a homogeneous
but anisotropic metric as

\begin{eqnarray}
	\label{eq1}
	ds^2=-dt^2+a(t)^{2}\,\sum_{i=1}^{3}\,e^{2\alpha_{i}(t)}\,(dx^{i})^{2}\,,
\end{eqnarray}
where the average of $\alpha_{i}(t)$ is defined as
$\bar{\alpha}(t)=\frac{1}{3}\Sigma_{i=1}^{3}\alpha_{i}(t)$. By this
background metric, we study the anisotropic tachyon inflation. We
show that when we consider the anisotropic metric, its effect
appears in the slow-roll parameters of the tachyon inflation. Since
the slow-roll parameters are included in the definition of the
perturbation parameters, they affect the observational viability of
the model. In this paper, we consider the
intermediate inflation in the tachyon anisotropic model. It should
be noticed that in Ref.~\cite{Cam09} the intermediate tachyon
inflation has been studied in detail. The authors of that paper, by
considering an intermediate scale factor, have obtained the
corresponding potential and studied the inflation and perturbations
in their setup. Interestingly, they have compared the behavior of
the tensor-to-scalar ratio versus the scalar spectral index, with
three years and five years WMAP data. Based on their study, the
intermediate tachyon inflation is consistent with WMAP data if the
value of the intermediate parameter is between 0 and 0.5. However,
it seems their model is not consistent with newly released data as
Planck2018 TT, TE, EE +lowE+lensing+BK14+BAO and Planck2018 TT, TE,
EE +lowE+lensing+BK18+BAO data. Note
that, from Planck2018 TT, TE, EE +lowE+lensing+BK14+BAO and based on
$\Lambda$CDM$+r+\frac{dn_{s}}{d\ln k}$ model, we have
$n_{s}=0.9658\pm0.0038$ and $r<0.072$~\cite{Pl18a,Pl18b}. A tighter
constraint has been implied on the tensor-to-scalar ratio by
Planck2018 TT, TE, EE +lowE+lensing+BK18+BAO data as $r<0.036$.
We wonder if considering the intermediate
tachyon model in an anisotropic geometry improves its observational
viability. In fact, we will show that the presence of anisotropic
inflation helps us to find some domain in parameters space that
makes the model consistent with new observational
data.

With these prerequisites, this paper is organized as follows. In
section \ref{sec2} we study the anisotropic tachyon inflation. In
section \ref{sec3}, by adopting the intermediate scalae factor, we
seek the observational viability of our setup. The reheating phase
after inflation is studied in section \ref{sec4}. Finally, we
present a summary of our study in section \ref{sec5}.

\section{\label{sec2}Anisotropic Tachyon Inflation}
To study the anisotropic inflation in the tachyon model, we use
metric (\ref{eq1}) as the background metric. As it has been
demonstrated in Ref.~\cite{Noj22}, if we redefine $a(t)\rightarrow
a(t)+\bar{\alpha}(t)$ and $\alpha_{i}(t)\rightarrow
\alpha_{i}(t)-\bar{\alpha}(t)$, we get
$\Sigma_{i=1}^{3}\alpha^{i}=0$ and
$\Sigma_{i=1}^{3}\dot{\alpha}^{i}=0$. Therefore, the components of
the Ricci tensor are given by~\cite{Noj22}
\begin{eqnarray}
	\label{eq2}
	R_{00}=-3\dot{H}-3H^{2}-\sum_{i=1}^{3}(\dot{\alpha}^{i})^2\,,
\end{eqnarray}
\begin{eqnarray}
	\label{eq3}
	R_{ij}=a^2e^{2\alpha_{i}}\left(\dot{H}+3H^2+\ddot{\alpha}^{i}+3H\dot{\alpha}^{i}\right)\delta_{ij}\,.
\end{eqnarray}
Note that, in the above equations and also the following ones, a dot
shows a time derivative of the parameter with respect to the cosmic
time. The Ricci scalar is obtained as
\begin{eqnarray}
	\label{eq4} R=6\dot{H}+12H^{2}+\sum_{i=1}^{3}(\dot{\alpha}^{i})^2\,.
\end{eqnarray}
By using the equations (\ref{eq2})-(\ref{eq4}) and
$T_{00}=-g_{00}\rho$ and $T_{ij}=-g_{ij}\,p$, we obtain the
Friedmann equations from the Einstein field equations
$G_{\mu\nu}=-\kappa^{2}\,T_{\mu\nu}$ as follows
\begin{eqnarray}
	\label{eq5}
	H^2=\frac{\kappa^{2}}{3}\rho+\frac{1}{6}\sum_{i=1}^{3}(\dot{\alpha}^{i})^{2}\,,
\end{eqnarray}
\begin{eqnarray}
	\label{eq6}
	2\dot{H}+3H^{2}=-\kappa^{2}\,p-\frac{1}{2}\sum_{i=1}^{3}(\dot{\alpha}^{i})^2\,.
\end{eqnarray}
Since we consider the tachyon field as an energy component of the
universe, the parameters $\rho$ and $p$ in equations (\ref{eq5}) and
(\ref{eq6}) are corresponding to this field and are given by
\begin{eqnarray}
	\label{eq7} \rho=\frac{V(\phi)}{\sqrt{1-\dot{\phi}^{2}}}\,,\quad
	p=-V(\phi)\,\sqrt{1-\dot{\phi}^{2}}\,.
\end{eqnarray}
However, the equation of motion of the tachyon field remains the
same and is given by
\begin{eqnarray}
	\label{eq8}\frac{\ddot{\phi}}{1-\dot{\phi}^{2}}+3\,H\dot{\phi}
	+\frac{V'(\phi)}{V(\phi)}=0\,,
\end{eqnarray}
where a prime shows the derivative of the parameter with respect to
the scalar field. Also, the sound speed is defined as
\begin{eqnarray}
	\label{eq9}c_{s}=\sqrt{1-\dot{\phi}^{2}}\,.
\end{eqnarray}
Now, from equations (\ref{eq5})-(\ref{eq7}), we find the following
expression for the potential
\begin{eqnarray}
	\label{eq10}V={\frac
		{6\,{H}^{2}-\sum_{i=1}^{3}(\dot{\alpha}^{i})^2}{2\,{\kappa}^{2}}\sqrt
		{{\frac {6\,{H}^{2}+4 \,{\it
						\dot{H}}+\sum_{i=1}^{3}(\dot{\alpha}^{i})^2}{6\,{H}^{2}-\sum_{i=1}^{3}(\dot{\alpha}^{i})^2}}}}
	\,,
\end{eqnarray}
and the time derivative of the tachyon field becomes as
\begin{eqnarray}
	\label{eq11}\dot{\phi}= \hspace{14cm}\nonumber\\{\frac {\sqrt
			{-36{H}^{4}+ \left( -12\sum_{i=1}^{3}(\dot{\alpha}^{i})^2-48\dot{H}
				\right) {
					H}^{2}+4{V}^{2}{\kappa}^{4}-{(\sum_{i=1}^{3}(\dot{\alpha}^{i})^2)}^{2}-8
				\dot{H}\sum_{i=1}^{3}(\dot{\alpha}^{i})^2-16{ \dot{H}}^{2
		}}}{2{\kappa}^{2}V}}\,,\nonumber\\
\end{eqnarray}
where the potential $V$ is given by equation (\ref{eq10}).
Therefore, the sound speed takes the following form
\begin{eqnarray}
	\label{eq12}c_{s}=\sqrt{1-{\frac {-36\,{H}^{4}+ \left( -12y-48
				\dot{H} \right) {H}^{2}+
				4\,{V}^{2}{\kappa}^{4}-{(\sum_{i=1}^{3}(\dot{\alpha}^{i})^2)}^{2}-8\dot{H}\sum_{i=1}^{3}(\dot{\alpha}^{i})^2-16\,{\dot{H}}^{2}}{4{V}^
				{2}{\kappa}^{4}}}
	}\,.\nonumber\\
\end{eqnarray}
For the tachyon field, the slow-roll parameters are defined
as~\cite{St04}
\begin{eqnarray}
	\label{eq13}\epsilon=\frac{1}{2\kappa^2}\frac{V'^{2}}{V^{3}}\,,
\end{eqnarray}
\begin{eqnarray}
	\label{eq14}\eta=\frac{1}{\kappa^2}\Big(-2\frac{V''}{V^{2}}+3\frac{V'^{2}}{V^{3}}\Big)\,,
\end{eqnarray}
and
\begin{eqnarray}
	\label{eq15}\zeta=\frac{1}{\kappa^2}\bigg(2\frac{V'''\,V'}{V^{4}}-10\frac{V''\,V'^{2}}{V^{5}}+9\frac{V'^{4}}{V^{6}}\bigg)\,.
\end{eqnarray}
Now, by using equations (\ref{eq10})-(\ref{eq12}), the first
slow-roll parameter takes the following form
\begin{eqnarray}
	\label{eq16}\epsilon=-\Bigg[  \left( -12 H^{2}
	+2\sum_{i=1}^{3}(\dot{\alpha}^{i})^2 \right) \ddot{H} -24H
	\dot{H}^{2}+ \left( -72 H^{3}+2{\frac {\rm d}{{\rm
				d}t}}(\sum_{i=1}^{3}(\dot{\alpha}^{i})^2) \right)\dot{H}
	\nonumber\\+\sum_{i=1}^{3}(\dot{\alpha}^{i})^2 {\frac {\rm d}{{\rm
				d}t}}(\sum_{i=1}^{3}(\dot{\alpha}^{i})^2) \Bigg]^{2} \Bigg[ \left( 6
	H^{2}+4 \dot{H} +\sum_{i=1}^{3}(\dot{\alpha}^{i})^{2} \right)^{5/2}
	\left( 6 H^{2}-\sum_{i=1}^{3}(\dot{\alpha}^{i})^{2} \right)^{3/2}\nonumber\\
	\left( 4\dot{H}+2\sum_{i=1}^{3}(\dot{\alpha}^{i})^{2}
	\right)\Bigg]^{-1} \,,
\end{eqnarray}
Since the obtained expressions for $\eta$
and $\zeta$ are very long equations, we present them in Appendix A
and Appendix B. In the absence of the anisotropy
property, $\alpha_{i}=0$, the slow-roll equations meet the
corresponding equations in the ordinary tachyon inflation. To
understand the effects of considering the anisotropic geometry on
the observational viability of the model, we can use the
perturbation parameters. The scalar
spectral index and the tensor-to-scalar ratio in the tachyon model
are defined, up to second order, as~\cite{St04}
\begin{eqnarray}
	\label{eq18}n_{s}-1=-2\epsilon-\eta-\bigg(2\epsilon^{2}+[2{\cal{F}}+3-2\chi]\epsilon\eta+{\cal{F}}\zeta\bigg)\,,
\end{eqnarray}
and
\begin{eqnarray}
	\label{eq19}r=16\epsilon\left(\,c_s+{\cal{F}\eta}\right)\,,
\end{eqnarray}
where, ${\cal{F}}\simeq-0.72$ and $\chi=\frac{1}{6}$. To find some
details about equations (\ref{eq18}) and (\ref{eq19}) see
Refs.~\cite{St04,Ave10,Cam09}. In the next section,
we study the viability of our setup in comparison with observational
data.

\section{\label{sec3}Observational Viability of the Model}

\begin{figure}[]
\begin{center}
\includegraphics[scale=0.3]{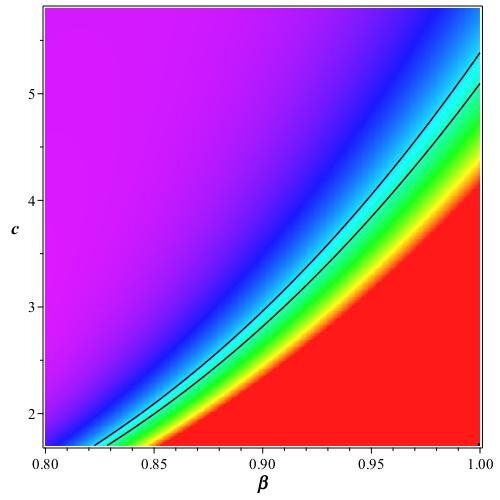}
\includegraphics[scale=0.32]{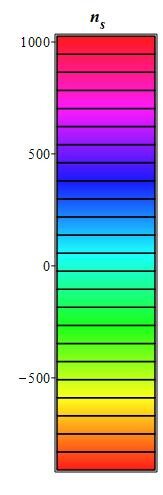}
\includegraphics[scale=0.3]{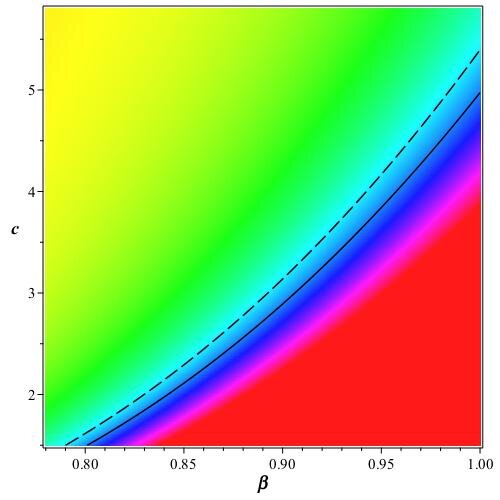}
\includegraphics[scale=0.32]{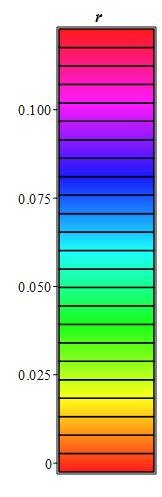}
\end{center}
\caption{\small {The prediction of the intermediate
		anisotropic tachyon inflation for the scalar spectral index and
		tensor-to-scalar ratio in some domains of the model's parameters
		space. In the left panel, the region between two solid lines leads
		to the observationally viable values of $n_{s}$, obtained from
		Planck2018 TT, TE, EE +lowE+lensing+BK14 +BAO data. In the right
		panel, the region above the solid line leads to the observationally
		viable values of $r$ obtained from Planck2018 TT, TE, EE
		+lowE+lensing+BK14 +BAO data. Also, the region above the dashed line
		leads to the observationally viable values of $r$ obtained from
		Planck2018 TT, TE, EE +lowE+lensing+BK18+BAO data. }}
\label{fig1}
\end{figure}

\begin{figure}
	\begin{center}
		\includegraphics[scale=0.12]{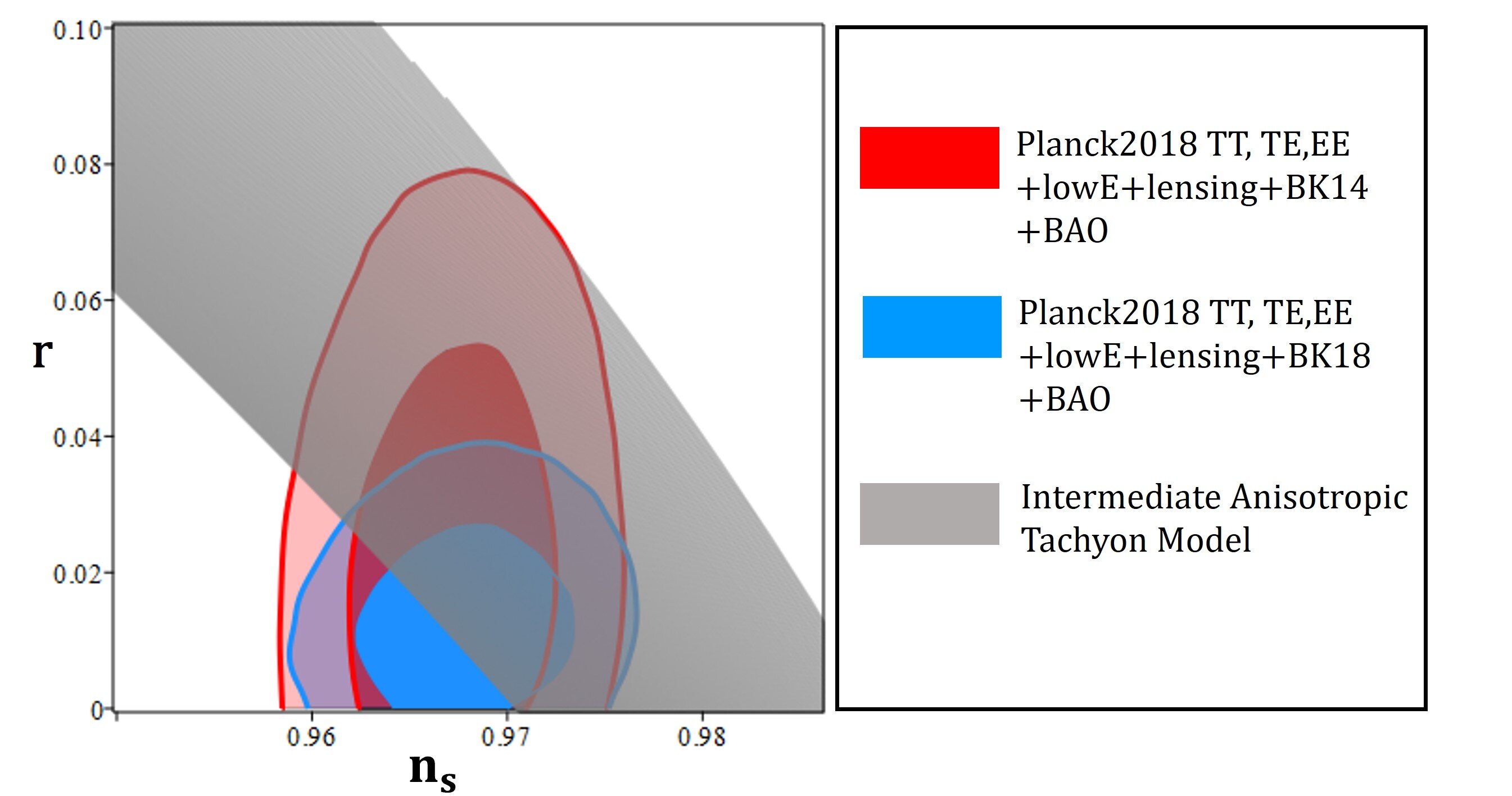}
	\end{center}
	\caption{\label{fig2}\small {Tensor-to-scalar ratio versus the
			scalar spectral index for the intermediate anisotropic tachyon
			inflation, in the background of Planck2018 TT, TE, EE
			+lowE+lensing+BK14 +BAO and Planck2018 TT, TE, EE +lowE+lensing+BK18
			+BAO data.}}
\end{figure}

\begin{figure}
	\begin{center}
			\includegraphics[scale=0.32]{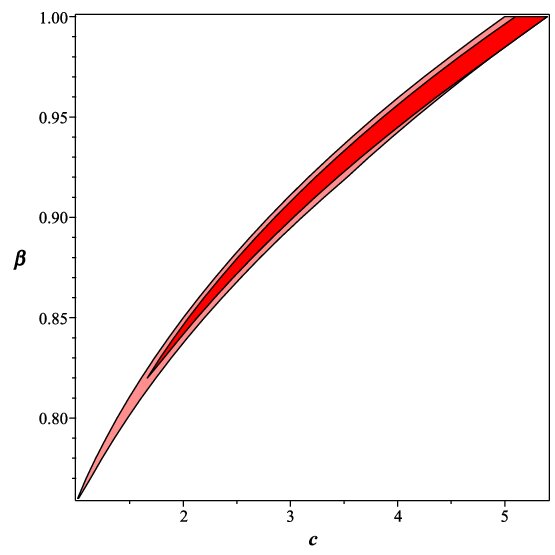}
				\includegraphics[scale=0.32]{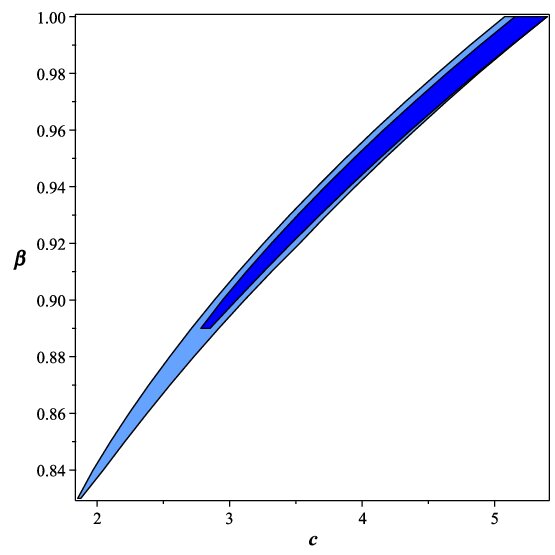}
	\end{center}
	\caption{\label{fig3}\small {Left panel: the parameters space of $c$
			and $\beta$ for the intermediate anisotropic tachyon inflation,
			leading to observationally viable values of $r-n_{s}$, in
			confrontation with Planck2018 TT, TE, and EE+lowE+lensing+BAO+BK14
			data at $68\%$ CL (dark red) and $95\%$ CL (light red). Right panel:
			the parameters space of $c$ and $\beta$ for the intermediate
			anisotropic tachyon inflation, leading to observationally viable
			values of $r-n_{s}$, in confrontation with Planck2018 TT, TE, and
			EE+lowE+lensing+BAO+BK18 data at $68\%$ CL (dark blue) and $95\%$ CL
			(light blue).}}
\end{figure}

When a new setup of inflation model is proposed, it is important to
examine its observational viability beyond its technical
calculations. Now that, we have obtained the main parameters of the
tachyon model in the anisotropic background, we perform some
numerical analysis on the model and try to constrain the model's
parameters observationally. To get this purpose, we need to first
identify some functions in our equations. It has been shown in
Ref.~\cite{Noj22} that $\alpha$ which is a function of time, satisfy
the following equation
\begin{eqnarray}
	\label{eq20}\ddot{\alpha}^{i}+3\,H\,\dot{\alpha}^{i}=0\,.
\end{eqnarray}
Solving this equation gives us the following expression for
$\dot{\alpha}^{i}$
\begin{eqnarray}
	\label{eq21}\dot{\alpha}^{i}=\frac{c^{i}}{a^{3}}\,,
\end{eqnarray}
where $c^{i}$'s are constant parameters. Also, since we have
$\Sigma_{i=1}^{3}\dot{\alpha}^{i}=0$, equation (\ref{eq21}) gives
the constraint $\Sigma_{i=1}^{3}c^{i}=0$. In the next step, we need
to adopt a scale factor. We are interested in working with
intermediate scale factor as
\begin{eqnarray}
	\label{eq22}a=a_{0}\,\exp (b\,t^{\beta})\,,
\end{eqnarray}
where $b$ is a constant and $0<\beta<1$. Now, by
using the definition of the number of e-folds during the inflation
as $N=\int H\, dt$ (with $H=\frac{\dot{a}}{a}$ being the Hubble
parameter), we obtain the first slow-roll parameter
in the intermediate anisotropic tachyon inflation as follows
\begin{eqnarray}
	\label{eq23}\epsilon=-2592\,{{\rm e}^{6\,N}}{\beta}^{2} \Bigg(
	\frac{1}{6}{{\rm e}^{6N}}{b}^{- 2{\beta}^{-1}}\beta\,{c}^{2} \left(
	\beta-1 \right) {N}^{{\frac {2\,
				\beta+2}{\beta}}}+\frac{{c}^{4}}{12}{N}^{{\frac
			{\beta+4}{\beta}}}{b}^{-4\,{\beta} ^{-1}}+  \left( \beta-1 \right)
	\bigg( -\frac{{b}^{-2{\beta}^ {-1}}}{36}\nonumber\\{c}^{2}{{\rm
			e}^{6N}} \left( \beta-2 \right) {N}^{{\frac {\beta
				+2}{\beta}}}+{\beta}^{2}{{\rm e}^{12\,N}}{N}^{3} \left(
	-\frac{2}{3}+ \left(
	N +\frac{1}{2} \right) \beta \right)  \bigg)  \Bigg) ^{2} \nonumber\\
	\left( 6\,{\beta}^{2 }{N}^{2}{{\rm
			e}^{6\,N}}-{c}^{2}{N}^{2\,{\beta}^{-1}}{b}^{-2\,{\beta}^ {-1}}
	\right) ^{-\frac{3}{2}} \left(
	{c}^{2}{N}^{2\,{\beta}^{-1}}{b}^{-2\,{ \beta}^{-1}}+6 \left(
	-\frac{2}{3}+ \left( N+\frac{2}{3} \right) \beta \right) { {\rm
			e}^{6\,N}}\beta\,N \right) ^{-\frac{5}{2}}\nonumber\\ \left(
	{c}^{2}{N}^{2\,{\beta}^ {-1}}{b}^{-2\,{\beta}^{-1}}+2\,\beta\,N{{\rm
			e}^{6\,N}} \left( \beta-1 \right)  \right) ^{-1} \,,
\end{eqnarray}
where, $c^{2}\equiv\sum_{i=1}^{3} {c^{i}}^{2}$. Again, we present
the slow-roll parameters $\eta$ and $\zeta$ in Appendix C and
Appendix D. By substituting the slow-roll parameters in equations
(\ref{eq18}) and (\ref{eq19}), we perform a numerical analysis in
our model to seek its viability in confrontation with the recent
observational data. The prediction of the intermediate anisotropic
tachyon inflation for the scalar spectral index and tensor-to-scalar
ratio, for $N=60$, in our model is shown in figure \ref{fig1}. As the figure
shows, there are some regions in the model's parameters space
leading to observationally viable values of $n_{s}$ and $r$. To show
the viability of the model more clearly, we have plotted $r-n_{s}$
behavior in the background of both Planck2018 TT, TE, EE
+lowE+lensing+BK14+BAO and Planck2018 TT, TE, EE
+lowE+lensing+BK18+BAO datasets in figure \ref{fig2}.
Our numerical analysis shows the
observational viability of the model for $0.82<\beta<1$ and
$1.641<c<5.375$ at $68\%$ CL and $0.76<\beta<1$ and $0.760<c<5.375$
at $95\%$ CL, in comparison with Planck2018 TT, TE, EE
+lowE+lensing+BK14+BAO data. Also, we have found the observational
viability of the model in comparison with Planck2018 TT, TE, EE
+lowE+lensing+BK18+BAO data for $0.89<\beta<1$ and $2.788<c<5.379$
at $68\%$ CL and $0.83<\beta<1$ and $1.854<c<5.379$ at $95\%$ CL.
For some sample values of $\beta$, we have summarized the
constraints on parameter $c$ in table \ref{tab1}. Note that, all these analysis have done for $N=60$. In figure
\ref{fig3}, we have plotted the observationally viable domains of
the $\beta -c$ parameters space, leading to the observationally
viable ranges of the parameters $n_{s}$ and $r$. To plot this
figure, we have used Planck2018 TT, TE, EE +lowE+lensing+BK14+BAO
and Planck2018 TT, TE, EE +lowE+lensing+BK18+BAO data at both $68\%$
and $95\%$ CL.

\begin{figure}
	\begin{center}
		\includegraphics[scale=0.32]{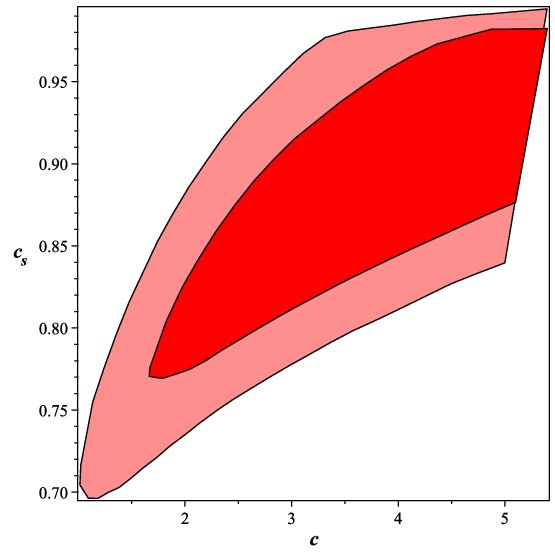}
		\includegraphics[scale=0.32]{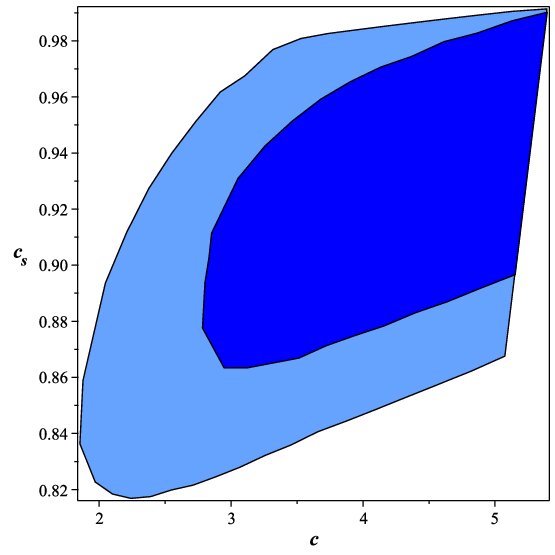}
		\end{center}
	\caption{\label{fig4}\small {Left panel: the parameter space of
			$c_{s}$ and $c$ for the intermediate anisotropic tachyon inflation,
			corresponding to observationally viable values of $r-n_{s}$,
			obtained from Planck2018 TT, TE, and EE+lowE+lensing+BAO+BK14 data
			at $68\%$ CL (dark red) and $95\%$ CL (light red). Right panel: the
			parameter space of $c_{s}$ and $c$ for the intermediate anisotropic
			tachyon inflation, corresponding to observationally viable values of
			$r-n_{s}$, obtained from Planck2018 TT, TE, and
			EE+lowE+lensing+BAO+BK18 data at $68\%$ CL (dark blue) and $95\%$ CL
			(light blue).}}
\end{figure}

\begin{figure}
	\begin{center}
		\includegraphics[scale=0.32]{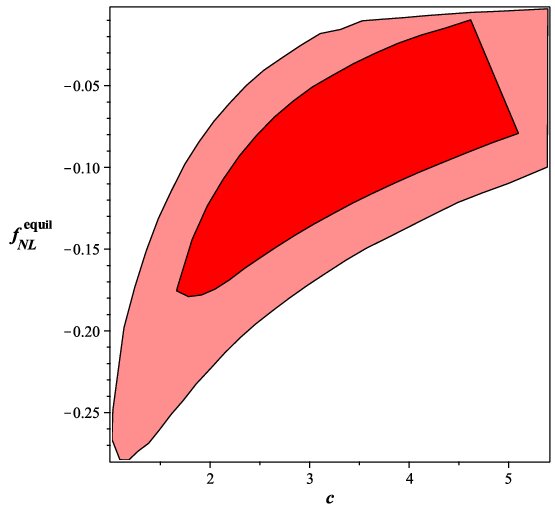}
		\includegraphics[scale=0.32]{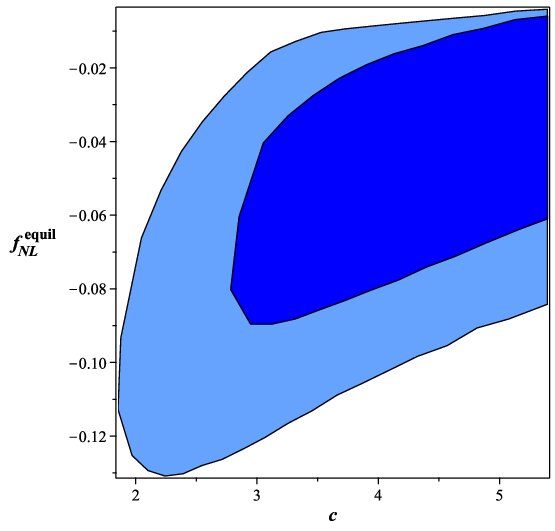}
	\end{center}
	\caption{\label{fig5}\small {Left panel: the parameter space of
			$f_{NL}^{equil}$ and $c$ for the intermediate anisotropic tachyon
			inflation, corresponding to observationally viable values of
			$c_{s}$, obtained from Planck2018 TT, TE, and
			EE+lowE+lensing+BAO+BK14 data at $68\%$ CL (dark red) and $95\%$ CL
			(light red). Right panel: the parameter space of $f_{NL}^{equil}$
			and $c$ for the intermediate anisotropic tachyon inflation,
			corresponding to observationally viable values of $c_{s}$, obtained
			from Planck2018 TT, TE, and EE+lowE+lensing+BAO+BK18 data at $68\%$
			CL (dark blue) and $95\%$ CL (light blue).}}
\end{figure}

Planck collaboration has also released some constraints on the
amplitude of the non-gausianity. The constraint on the equilateral
amplitude of the non-gaussianity, obtained from Planck2018 TTT, EEE,
TTE, and EET data, is $f_{NL}^{equil}=-26\pm 47$~\cite{Pl19}.  Now,
we consider the following relation between the sound speed and the
equilateral amplitude of the non-gaussianity
\begin{eqnarray}
	\label{eq25}f_{NL}^{equil}=\frac{85}{324}\Bigg(1-\frac{1}{c_{s}^{2}}\Bigg)\,.
\end{eqnarray}
From equations (\ref{eq12}) and (\ref{eq25}), and by using the
observational constraint on the $f_{NL}^{equil}$, we have found that
for all values of $0<\beta< 1$ and $c>0$  the amplitude of the
non-gaussianity in the intermediate anisotropic tachyon inflation is
consistent with Planck2018 TTT, EEE, TTE, and EET data. However, we
are interested in those values of $c$ and $\beta$ that give
observationally viable values of all $n_{s}$, $r$, and
$f_{NL}^{equil}$. In this regard, we consider the observationally
viable ranges of $c$ and $\beta$, which have been obtained from the
viable ranges of $r-n_{s}$ in confrontation with Planck2018 TT, TE,
and EE+lowE+lensing+BAO+BK14(18) data. By using these constraints,
we have plotted the parameter space $c_{s}-c $ in figure \ref{fig4}.
We have also obtained some constraints on the sound speed, for some
sample values of $\beta$, that have been summarized in table
\ref{tab2}. These constraints on $c$ and $c_{s}$, give us some
information about the amplitude of the non-gaussianity. In figure
\ref{fig5} we have plotted the parameter space $f_{NL}^{equil}-c$.
This figure shows that for those domains of beta, that lead to the
observationally viable values of $r-n_{s}$, the amplitude of the
non-gaussianity is ${\cal{O}}(10^{-2}-10^{-1})$. We have also
summarized some constraints on the equilateral amplitude of the
non-gaussianity in table \ref{tab2}.

\begin{table*}
	\tiny\tiny\caption{\small{\label{tab1} Ranges of the parameter $c$,
			for some sample values of $\beta$, in which the tensor-to-scalar
			ratio and scalar spectral index of the intermediate anisotropic
			tachyon model are consistent with different data sets.}}
	\begin{center}
		\tabcolsep=0.05cm\begin{tabular}{cccccc}
			\\ \hline \hline \\ & Planck2018 TT,TE,EE+lowE & Planck2018 TT,TE,EE+lowE&Planck2018 TT,TE,EE+lowE&Planck2018 TT,TE,EE+lowE
			\\
			& +lensing+BK14+BAO &
			+lensing+BK14+BAO&lensing+BK18+BAO&lensing+BK18+BAO
			\\
			\hline \\$\beta$& $68\%$ CL & $95\%$ CL &$68\%$ CL & $95\%$ CL
			\\
			\hline\hline \\  $0.80$ & Not consistent & $1.382<c<1.480$
			& Not consistent & Not consistent\\ \\
			\hline
			\\$0.85$& $2.049<c<2.129$ & $2.00<c<2.197$
			&Not consistent& $2.101<c<2.209$
			\\ \\ \hline\\
			$0.90$& $2.855<c<3.028$ & $2.796<c<3.107
			$ & $2.947<c<3.052$ & $2.885<c<3.113$ \\ \\
			\hline\\
			$0.95$& $3.864<c<4.122$ & $3.788<c<4.169$ & $3.937<c<4.132$ &
			$3.873<c<4.169$
			\\ \\
			\hline \hline
		\end{tabular}
	\end{center}
\end{table*}

\begin{table*}
	\tiny\tiny\caption{\small{\label{tab2} Ranges of the observationally
			viable values of the sound speed and corresponding equilateral
			amplitude of the non-gaussianity, in the intermediate anisotropic
			tachyon model and in confrontation with different data sets.}}
	\begin{center}
		\tabcolsep=0.05cm\begin{tabular}{ccccccccc}
			\\ \hline \hline \\ & Planck2018 TT,TE,EE+lowE && Planck2018 TT,TE,EE+lowE && Planck2018 TT,TE,EE+lowE && Planck2018 TT,TE,EE+lowE
			\\
			& +lensing+BK14+BAO &&
			+lensing+BK14+BAO&&lensing+BK18+BAO&&lensing+BK18+BAO
			\\
			\hline \\$\beta$& $68\%$ CL && $95\%$ CL &&$68\%$ CL && $95\%$ CL
			\\
			\hline\hline \\  $0.80$ & Not consistent && $0.159<c_{s}<0.185$ &&
			Not consistent && Not consistent
			\\ \\  & Not consistent && $-0.268<f_{NL}<-0.131$
			&& Not consistent && Not consistent\\ \\
			\hline
			\\$0.85$& $0.176<c_{s}<0.191$ && $0.167<c_{s}<0.204$
			&&Not consistent&& $0.185<c_{s}<0.207$
			\\\\& $-0.174<f_{NL}^{equil}<-0.107$ &&
			$-0.223<f_{NL}^{equil}<-0.060$ &&Not consistent&&
			$-0.129<f_{NL}^{equil}<-0.053$
			\\ \\ \hline\\
			$0.90$& $0.183<c_{s}<0.208$ && $0.175<c_{s}<0.219 $ &&
			$0.196<c_{s}<0.211$ && $0.187<c_{s}<0.223$
			\\\\
			& $-0.141<f_{NL}^{equil}<-0.050$ && $-0.179<f_{NL}^{equil}<-0.018
			$ && $-0.089<f_{NL}^{equil}<-0.040$ && $-0.123<f_{NL}^{equil}<-0.008$ \\ \\
			\hline\\
			$0.95$& $0.190<c_{s}<0.219$ && $0.176<c_{s}<0.222$ &&
			$0.198<c_{s}<0.220$ && $0.188<c_{s}<0.220$
			\\\\
			& $-0.109<f_{NL}^{equil}<-0.019$& & $-0.142<f_{NL}^{equil}<-0.007$
			&& $-0.080<f_{NL}^{equil}<-0.016$ && $-0.105<f_{NL}^{equil}<-0.007$
			\\ \\
			\hline \hline
		\end{tabular}
	\end{center}
\end{table*}

\section{\label{sec4}Reheating}

When the initial inflation expansion of the universe ends, there
should be a phase that by reheating the universe makes it ready for
the subsequent evolution. This reheating phase is one important
process in the history of the universe and studying its cosmology
gives us some interesting information. It is
believed that when inflation terminates, the inflaton rolls down to
the minimum of its potential and starts oscillating there. The
oscillation causes the field to lose its energy and decay to
relativistic particles~\cite{Far82,Alb82}. Considering the fact that
at the beginning of inflation, the effective equation of state
parameter is equal to -1 and at the end of this era it reaches
$-\frac{1}{3}$, it is important to know how the universe should be
prepared for the subsequent eras. This would be more important when
we consider that the value of the equation of state at the beginning
of the radiation dominated phase is $\frac{1}{3}$. It has been shown
in Ref.~\cite{Pod06} that, in the case of massive inflation, the
field initially oscillates at a rate much larger than the expansion
rate. This, with a good approximation, leads to a zero effective
equation of state at the beginning of the reheating phase. By more
oscillating of the field and decaying to other particles, the value
of the effective equation of state parameter changes, and at the end
of the reheating phase it becomes equal to $\frac{1}{3}$. There are
also some other scenarios such as instant reheating~\cite{Fel99},
tachyon instability~\cite{Gre97,Shu06,Duf06} and resonance
decay~\cite{Tra90,Kof94,Kof97}, where the reheating occurs in a
non-perturbative process. To study the
reheating phase in the intermediate anisotropic tachyon model, we
follow Refs.~\cite{Dai14,Un15,Co15,Cai15,Ue16} and find the number
of e-folds during the reheating phase ($N_{rh}$) in our model. By
starting the definition of the e-folds number as
\begin{equation}
	\label{eq26} N=\ln \left(\frac{a_{e}}{a_{hc}}\right)\,,
\end{equation}
that has been defined between the Hubble crossing ($hc$) of the
physical scales and the end ($e$) of inflation, we try to find some
expression for $N_{rh}$ in terms of the model's parameters. Since
the scale factor is related to the energy density as $\rho\sim
a^{-3(1+\omega_{eff})}$, where $\omega_{eff}$ is the effective
equation of state during the reheating, we have
\begin{eqnarray}
	\label{eq27}
	N_{rh}=\ln\left(\frac{a_{rh}}{a_{e}}\right)=-\frac{1}{3(1+\omega_{eff})}\ln\left(\frac{\rho_{rh}}{\rho_{e}}\right)\,.
\end{eqnarray}
Now, we use the relation between the wave number and the scale
factor at the horizon crossing as
\begin{eqnarray}
	\label{eq28} 0=\ln\left(\frac{k_{hc}}{a_{hc}H_{hc}}\right)=
	\ln\left(\frac{a_{e}}{a_{hc}}\frac{a_{rh}}{a_{e}}\frac{a_{0}}{a_{rh}}\frac{k_{hc}}{a_{0}H_{hc}}\right)\,,
\end{eqnarray}
where the subscript ``$0$'' stands for the current value of the
corresponding parameter, we get
\begin{eqnarray}
	\label{eq29}
	N+N_{rh}+\ln\left(\frac{k_{hc}}{a_{0}H_{hc}}\right)+\ln\left(\frac{a_{0}}{a_{rh}}\right)=0\,.
\end{eqnarray}
If we consider the following relation between the energy density and
temperature~\cite{Co15,Ue16}
\begin{equation}
	\label{eq30} \rho_{rh}=\frac{\pi^{2}g_{rh}}{30}T_{rh}^{4}\,,
\end{equation}
with $g_{rh}$ being the effective number of the relativistic species
at the reheating phase, and also~\cite{Co15,Ue16}
\begin{equation}
	\label{eq31}
	\frac{a_{0}}{a_{rh}}=\left(\frac{43}{11g_{rh}}\right)^{-\frac{1}{3}}\frac{T_{rh}}{T_{0}}\,,
\end{equation}
we obtain
\begin{eqnarray}
	\label{eq32}
	\frac{a_{0}}{a_{rh}}=\left(\frac{43}{11g_{rh}}\right)^{-\frac{1}{3}}T_{0}^{-1}\left(\frac{\pi^{2}g_{rh}}{30\rho_{rh}}\right)^{-\frac{1}{4}}\,.
\end{eqnarray}
In our intermediate anisotropic tachyon model, we can write the
energy density as
\begin{eqnarray}
	\label{eq33}
	\rho=\frac{V\sqrt{18\kappa^{2}\,V+\sum_{i=1}^{3}(\dot{\alpha}^{i})^{2}}}{\sqrt{(18-12\epsilon)\kappa^{2}\,V-\sum_{i=1}^{3}(\dot{\alpha}^{i})^{2}}}\,.
\end{eqnarray}
Therefore, at the end of inflation, we have
\begin{eqnarray}
	\label{eq34}
	\rho_{e}=\frac{V_{e}\sqrt{18\kappa^{2}\,V_{e}+\sum_{i=1}^{3}(\dot{\alpha}^{i})^{2}}}
	{\sqrt{6\kappa^{2}\,V_{e}-\sum_{i=1}^{3}(\dot{\alpha}^{i})^{2}}}\,,
\end{eqnarray}
which by using equation (\ref{eq27}) leads to
\begin{eqnarray}
	\label{eq35}
	\rho_{rh}=\Bigg[\frac{V_{e}\sqrt{18\kappa^{2}\,V_{e}+\sum_{i=1}^{3}(\dot{\alpha}^{i})^{2}}}
	{\sqrt{6\kappa^{2}\,V_{e}-\sum_{i=1}^{3}(\dot{\alpha}^{i})^{2}}}\,\Bigg]\,\exp\Big[-3N_{rh}(1+\omega_{eff})\Big].
\end{eqnarray}
From the above equation and also equation (\ref{eq32}) we find
\begin{eqnarray}
	\label{eq36}
	\ln\left(\frac{a_{0}}{a_{rh}}\right)=-\frac{1}{3}\ln\left(\frac{43}{11g_{rh}}\right)
	-\frac{1}{4}\ln\left(\frac{\pi^{2}g_{rh}}{30\rho_{rh}}\right)-\ln
	T_{0}
	+\frac{1}{4}\ln\Bigg(\frac{V_{e}\sqrt{18\kappa^{2}\,V_{e}+\sum_{i=1}^{3}(\dot{\alpha}^{i})^{2}}}
	{\sqrt{6\kappa^{2}\,V_{e}-\sum_{i=1}^{3}(\dot{\alpha}^{i})^{2}}}\Bigg)\nonumber\\
	-\frac{3}{4}N_{rh}(1+\omega_{eff})\,.
\end{eqnarray}
Now, we find the number of e-folds during the reheating from
equations (\ref{eq29}) and (\ref{eq36}) as
\begin{eqnarray}
	\label{eq37}
	N_{rh}=\frac{4}{1-3\omega_{eff}}\Bigg[-N-\ln\Big(\frac{k_{hc}}{a_{0}T_{0}}\Big)-\frac{1}{4}\ln\Big(\frac{40}{\pi^{2}g_{rh}}\Big)
	-\frac{1}{3}\ln\Big(\frac{11g_{rh}}{43}\Big)\nonumber\\+\frac{1}{2}\ln\Big(H^{2}\Big)
	-\frac{1}{4}\ln\Bigg(\frac{V_{e}\sqrt{18\kappa^{2}\,V_{e}+\sum_{i=1}^{3}(\dot{\alpha}^{i})^{2}}}
	{\sqrt{6\kappa^{2}\,V_{e}-\sum_{i=1}^{3}(\dot{\alpha}^{i})^{2}}}\Bigg)\Bigg].
\end{eqnarray}
Also, from equations (\ref{eq27}), (\ref{eq31}), and (\ref{eq34}) we
find the temperature during the reheating phase as
\begin{equation}
	\label{eq38}
	T_{rh}=\bigg(\frac{30}{\pi^{2}g_{rh}}\bigg)^{\frac{1}{4}}\,
	\Bigg[\frac{V_{e}\sqrt{18\kappa^{2}\,V_{e}+\sum_{i=1}^{3}(\dot{\alpha}^{i})^{2}}}
	{\sqrt{6\kappa^{2}\,V_{e}-\sum_{i=1}^{3}(\dot{\alpha}^{i})^{2}}}\,\Bigg]^{\frac{1}{4}}\,\exp\bigg[-\frac{3}{4}N_{rh}(1+\omega_{eff})\bigg]\,.
\end{equation}

\begin{figure}
	\begin{center}
		\includegraphics[scale=0.32]{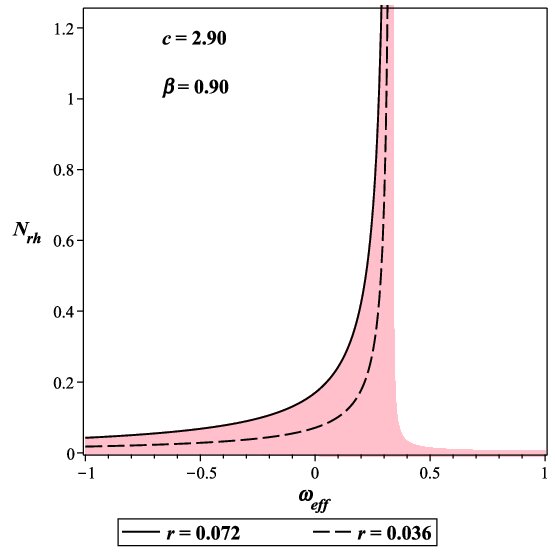}
		\includegraphics[scale=0.32]{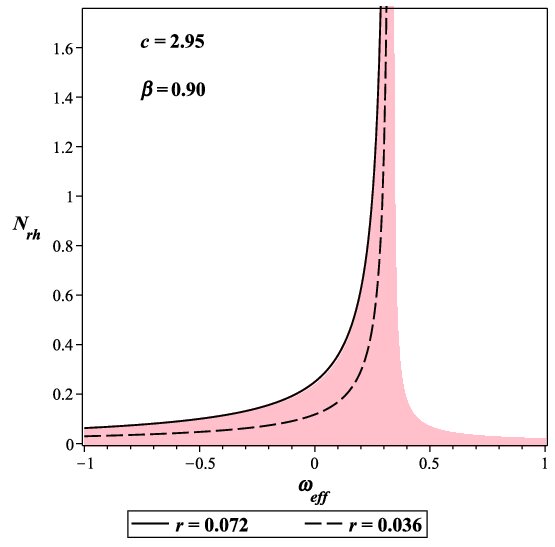}
	\end{center}
	\caption{\label{fig6}\small {Ranges of the e-folds number and the
			effective equation of state parameter during the reheating phase for
			the intermediate anisotropic tachyon inflation, leading to the
			observationally viable values of the tensor-to-scalar ratio.}}
\end{figure}

\begin{figure}
	\begin{center}
		\includegraphics[scale=0.32]{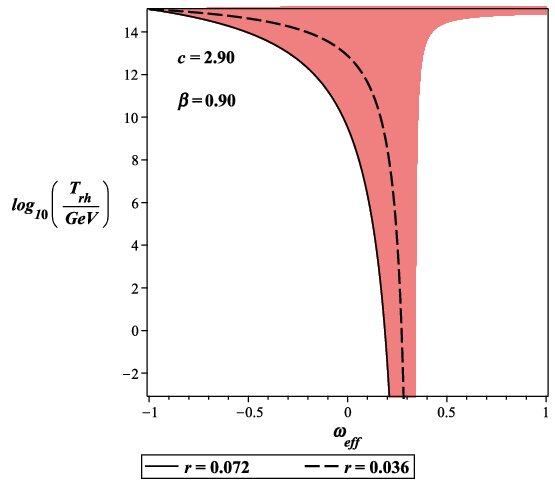}
		\includegraphics[scale=0.32]{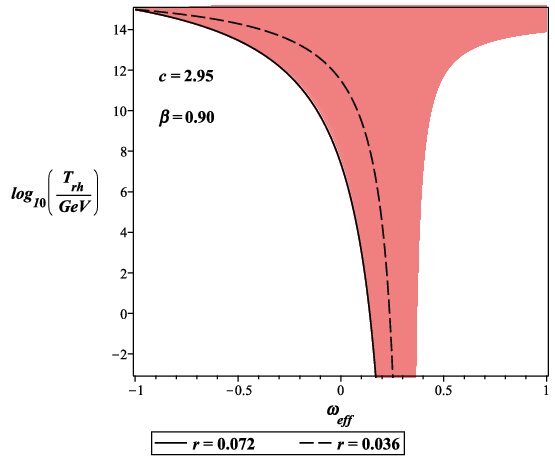}
		\end{center}
	\caption{\label{fig7}\small {Ranges of the temperature and the
			effective equation of state parameter during the reheating phase for
			the intermediate anisotropic tachyon inflation, leading to the
			observationally viable values of the tensor-to-scalar ratio.}}
\end{figure}

\begin{figure}
	\begin{center}
		\includegraphics[scale=0.32]{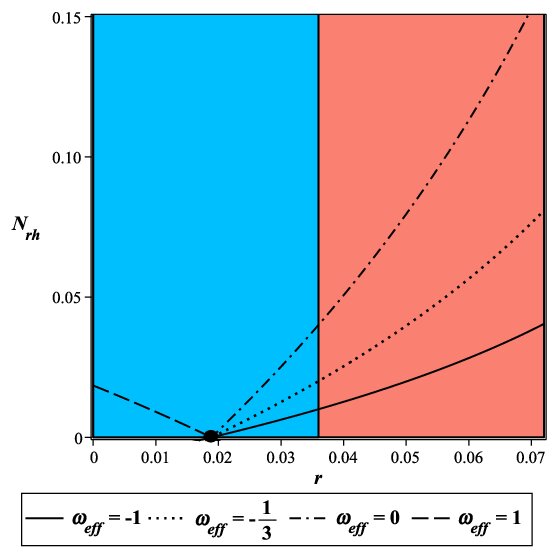}
		\includegraphics[scale=0.34]{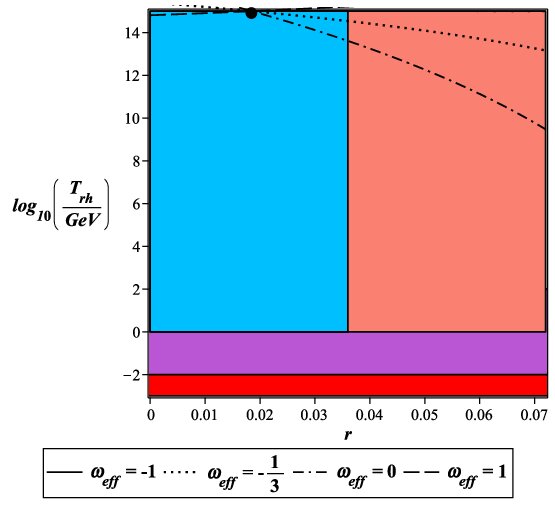}
	\end{center}
	\caption{\label{fig8}\small {Behavior of the e-folds number and
			temperature during the reheating phase versus the tensor-to-scalar
			ratio. The salmon and deep sky blue colors show the ranges of the
			observationally viable values of the tensor-to-scalar ratio,
			obtained from Planck2018 TT, TE, and EE+lowE+lensing+BAO+BK14 and
			Planck2018 TT, TE, and EE+lowE+lensing+BAO+BK18 data, respectively.
			Note that the salmon region overlaps with some domain of deep sky
			blue region. In the right panel, the purple region corresponds to
			the temperatures below the electroweak scale, $T<100$ GeV, and the
			red region corresponds to the temperatures below the big bang
			nucleosynthesis scale, $T<10$ MeV.}}
\end{figure}

To perform a numerical analysis on the reheating parameters, we use
the equation (\ref{eq10}) and also the general definition of the
slow-roll parameter as $\epsilon=-\frac{\dot{H}}{H^{2}}$ to find
equation (\ref{eq33}) in terms of $\epsilon$. Then, we consider
$\epsilon=1$ (corresponding to the end of the inflation) and find
equation (\ref{eq34}) in terms of the model's parameters. Also, from
equation (\ref{eq19}), we can obtain $\epsilon$ and therefore
$H^{2}$ in terms of the tensor-to-scalar ratio and substitute it in
equation (\ref{eq37}). In this way, we can perform some numerical
analysis in our model to study the reheating process. The
observational constraints on $r$ lead to some constraints on the
reheating parameters. We use both Planck2018 TT, TE, and
EE+lowE+lensing+BAO+BK14 and Planck2018 TT, TE, and
EE+lowE+lensing+BAO+BK18 constraints on the tensor-to-scalar ratio
as $r<0.072$ and $r<0.036$, respectively. Figure \ref{fig6} shows
the parameter space of $N_{rh}$ and $\omega_{eff}$ leading to the
observationally viable values of the tensor-to-scalar ratio.
In this figure, we have adopted two
sample values of $c$. As the figure shows, in this model it is
possible to have instantaneous reheating where $N_{rh}\rightarrow
0$. We have obtained some constraints on
$N_{rh}$, for some sample values of $c$ and $\beta$ that have been
summarized in table \ref{tab3}. Also, for consistency with Big-Bang
nucleosynthesis, the temperature during the reheating phase should
be larger or of the order of $10 MeV$. To check this consistency in
our model, we have performed some numerical analysis on $T_{rh}$.
The result is shown in figure \ref{fig7}, for two sample values of
$c$. It seems that this consistency is satisfied in the intermediate
anisotropic tachyon inflation. We have also obtained some
constraints on the reheating temperature that have been presented in
table \ref{tab4}. To show the
results of this section more clearly, we have plotted $N_{rh}-r$ and
$\log_{10}\Big(\frac{T_{rh}}{GeV}\Big)-r$ for some adopted values of
the equation of state parameter as
$\omega_{eff}=-1,-\frac{1}{3},0,1$. The results have been shown in
figure \ref{fig8}. This figure shows that all plotted curves meet at
one point where $\omega=\frac{1}{3}$ would pass vertically and also
it is observationally viable. In fact, this figure shows that
instantaneous reheating is possible, at least in some domains of the
parameter space. In Ref~\cite{Nau18} also, the possibility of having
instantaneous reheating in the tachyon model has been
shown. According to our study, the
intermediate anisotropic inflation gives the viable reheating
process after inflation.

\begin{table*}
	\tiny \tiny \caption{\small{\label{tab3} Constraints on the e-folds
			number during reheating in the intermediate anisotropic tachyon
			model, obtained from Planck2018 TT, TE, EE+lowE+lensing+BK14
			(18)+BAO joint data at $95\%$ CL.}}
	\begin{center}
		\begin{tabular}{ccccccc}
			\\ \hline \hline \\ &$\beta$&$c$& $\omega=-1$& $\omega=-\frac{1}{3}$
			&$\omega=0$
			\\
			\hline
			\\Planck2018 TT, TE, EE+&$0.90$&2.90&$0\leq N_{rh}\leq0.040$ &$0\leq N_{rh}\leq0.081$ &$0\leq N_{rh}\leq0.171$\\ \\
			lowE+lensing+BK14&$0.90$&2.95&$0\leq N_{rh}\leq0.058$&$0\leq N_{rh}\leq0.124$&$0\leq N_{rh}\leq0.250$\\ \\
			+BAO&$0.90$&3.00&$0\leq N_{rh}\leq0.075$&$0\leq N_{rh}\leq0.145$&$0\leq N_{rh}\leq0.295$\\ \\
			&$0.95$&3.90&$0\leq N_{rh}\leq0.079$&$0\leq N_{rh}\leq0.151$&$0\leq N_{rh}\leq0.301$\\ \\
			&$0.95$&3.95&$0\leq N_{rh}\leq0.085$&$0\leq N_{rh}\leq0.157$&$0\leq N_{rh}\leq0.307$\\ \\
			&$0.95$&4.00&$0\leq N_{rh}\leq0.091$&$0\leq N_{rh}\leq0.161$&$0\leq N_{rh}\leq 0.312$\\ \\
			\hline\\
			Planck2018 TT, TE, EE+&$0.90$&2.95&$0\leq N_{rh}\leq0.010$&$0\leq N_{rh}\leq0.035$&$0\leq N_{rh}\leq0.039$\\ \\
			lowE+lensing+BK18&$0.90$&3.00&$0\leq N_{rh}\leq0.026$&$0\leq N_{rh}\leq0.058$&$0\leq N_{rh}\leq0.113$\\ \\
			+BAO&$0.90$&3.05&$0\leq N_{rh}\leq0.064$&$0\leq N_{rh}\leq0.115$&$0\leq N_{rh}\leq0.216$\\ \\
			&$0.95$&3.90&$0\leq N_{rh}\leq0.073$&$0\leq N_{rh}\leq0.126$&$0\leq N_{rh}\leq0.230$\\ \\
			&$0.95$&3.95&$0\leq N_{rh}\leq0.077$&$0\leq N_{rh}\leq0.131$&$0\leq N_{rh}\leq0.235$\\ \\
			&$0.95$&4.00&$0\leq N_{rh}\leq0.083$&$0\leq N_{rh}\leq0.137$&$0\leq N_{rh}\leq0.242$\ \\
			\hline \hline
		\end{tabular}
	\end{center}
\end{table*}

\begin{table*}
	\tiny \tiny \caption{\small{\label{tab4} Constraints on the
			temperature during reheating in the intermediate anisotropic tachyon
			model, obtained from Planck2018 TT, TE, EE+lowE+lensing+BK14
			(18)+BAO joint data at $95\%$ CL.}}
	\begin{center}
		\begin{tabular}{ccccccc}
			\\ \hline \hline \\ &$\beta$&$c$& $\omega=-1$& $\omega=-\frac{1}{3}$
			&$\omega=0$
			\\
			\hline
			\\Planck2018 TT, TE, EE+&$0.90$&2.90&$14.99\leq \log_{10}\left(\frac{T_{rh}}{GeV}\right)\leq15.00$&
			$13.17\leq \log_{10}\left(\frac{T_{rh}}{GeV}\right)\leq15.00$&$9.46\leq \log_{10}\left(\frac{T_{rh}}{GeV}\right)\leq15.00$\\ \\
			lowE+lensing+BK14&$0.90$&2.95&$14.98\leq
			\log_{10}\left(\frac{T_{rh}}{GeV}\right)\leq15.00$
			&$12.43\leq \log_{10}\left(\frac{T_{rh}}{GeV}\right)\leq15.00$&$7.28\leq \log_{10}\left(\frac{T_{rh}}{GeV}\right)\leq15.00$\\ \\
			+BAO&$0.90$&3.00&$14.98\leq
			\log_{10}\left(\frac{T_{rh}}{GeV}\right)\leq15.00$&
			$11.65\leq \log_{10}\left(\frac{T_{rh}}{GeV}\right)\leq15.00$&$6.03\leq \log_{10}\left(\frac{T_{rh}}{GeV}\right)\leq15.00$\\ \\
			&$0.95$&3.90&$14.97\leq
			\log_{10}\left(\frac{T_{rh}}{GeV}\right)\leq15.00$&
			$10.93\leq \log_{10}\left(\frac{T_{rh}}{GeV}\right)\leq15.00$&$5.12\leq \log_{10}\left(\frac{T_{rh}}{GeV}\right)\leq15.00$\\ \\
			&$0.95$&3.95&$14.96\leq
			\log_{10}\left(\frac{T_{rh}}{GeV}\right)\leq15.00$&
			$10.47\leq \log_{10}\left(\frac{T_{rh}}{GeV}\right)\leq15.00$&$4.05\leq \log_{10}\left(\frac{T_{rh}}{GeV}\right)\leq15.00$\\ \\
			&$0.95$&4.00&$14.95\leq
			\log_{10}\left(\frac{T_{rh}}{GeV}\right)\leq15.00$&
			$10.00\leq \log_{10}\left(\frac{T_{rh}}{GeV}\right)\leq15.00$&$2.14\leq \log_{10}\left(\frac{T_{rh}}{GeV}\right)\leq15.00$\\ \\
			\hline\\
			Planck2018 TT, TE, EE+&$0.90$&2.90&$14.99\leq
			\log_{10}\left(\frac{T_{rh}}{GeV}\right)\leq15.00$&
			$14.28\leq \log_{10}\left(\frac{T_{rh}}{GeV}\right)\leq15.00$&$13.56\leq \log_{10}\left(\frac{T_{rh}}{GeV}\right)\leq15.00$\\ \\
			lowE+lensing+BK18&$0.90$&2.95&$14.99\leq
			\log_{10}\left(\frac{T_{rh}}{GeV}\right)\leq15.00$
			&$13.77\leq \log_{10}\left(\frac{T_{rh}}{GeV}\right)\leq15.00$&$11.50\leq \log_{10}\left(\frac{T_{rh}}{GeV}\right)\leq15.00$\\ \\
			+BAO&$0.90$&3.00&$14.91\leq
			\log_{10}\left(\frac{T_{rh}}{GeV}\right)\leq15.00$&
			$13.15\leq \log_{10}\left(\frac{T_{rh}}{GeV}\right)\leq15.00$&$10.68\leq \log_{10}\left(\frac{T_{rh}}{GeV}\right)\leq15.00$\\ \\
			&$0.95$&3.90&$14.95\leq
			\log_{10}\left(\frac{T_{rh}}{GeV}\right)\leq15.00$&
			$12.78\leq \log_{10}\left(\frac{T_{rh}}{GeV}\right)\leq15.00$&$9.74\leq \log_{10}\left(\frac{T_{rh}}{GeV}\right)\leq15.00$\\ \\
			&$0.95$&3.95&$14.85\leq
			\log_{10}\left(\frac{T_{rh}}{GeV}\right)\leq14.93$&
			$12.33\leq \log_{10}\left(\frac{T_{rh}}{GeV}\right)\leq15.00$&$9.01\leq \log_{10}\left(\frac{T_{rh}}{GeV}\right)\leq15.00$\\ \\
			&$0.95$&4.00&$14.9\leq
			\log_{10}\left(\frac{T_{rh}}{GeV}\right)\leq15.00$&
			$11.73\leq \log_{10}\left(\frac{T_{rh}}{GeV}\right)\leq15.00$&$8.24\leq \log_{10}\left(\frac{T_{rh}}{GeV}\right)\leq15.00$\\ \\
			\hline \hline
		\end{tabular}
	\end{center}
\end{table*}

\section{\label{sec5}Conclusion}
In this paper, we have studied the intermediate tachyon inflation in
the anisotropic geometry. We have obtained the Friedmann equations
in this anisotropic background. Since the Friedmann equations are
modified, the effects of the anisotropy are presented in the
potential of the model. In this regard, by using the modified
potential, we have obtained the slow-roll parameters in the
anisotropic tachyon inflation. It has been shown that the
anisotropic background changes the slow-roll parameters in the
tachyon model. Therefore, the important perturbation parameters,
such as the scalar spectral index and the tensor-to-scalar ratio are
changed too. This may change the viability of the model. To seek the
observational viability of this model, we have considered the
intermediate scale factor and obtained the slow-roll parameters in
the intermediate anisotropic tachyon inflation. We have analyzed the
model numerically and found the prediction of our model for the
scalar spectral index and tensor-to-scalar ratio. We have shown
that, in some domains of the model's parameters, the model is
observationally viable. To show the viability of the intermediate
anisotropic tachyon inflation more clearly, we have plotted the
$r-n_{s}$ behavior in the background of Planck2018 TT, TE,
EE+lowE+lensing+BK14(18)+BAO data. We have also obtained some
constraints on the parameter $c$ that have been summarized in
tables. By using the observational constraints, we have plotted the
$c-\beta$ parameters space leading to the observationally viable
values of $n_{s}$ and $r$, in confrontation with Planck2018 TT, TE,
EE+lowE+lensing+BK14(18)+BAO data at $68\%$ and $98\%$ CL. Given
that the tensor-to-scalar ratio and sound speed are related, by
using the constraints on $r$, we have found the observationally
viable ranges of the sound speed. In this way, we have also obtained
some constraints on the amplitude of the non-gaussianity in our
model. We have shown that the prediction of our model for the
equilateral amplitude of the non-gaussianity is of the order of
$10^{-2}-10^{-1}$. Finally, we have studied the reheating process
after inflation in our setup. By analyzing the number of e-folds and
temperature during the reheating, we have obtained some additional
information about the model. We have used the observational
constraints on the tensor-to-scalar ratio and obtained the viable
domains of $N_{rh}-\omega_{eff}$ and $T_{rh}-\omega_{eff}$
parameters space. We have shown in the intermediate anisotropic
tachyon inflation, it is possible to have instantaneous reheating.
Also, the temperature during the reheating is
consistent with Big-Bang nucleosynthesis.\\\\

\textbf{ACKNOWLEDGMENTS}\\
I thank the referee for the very insightful comments that have
improved the quality of the paper
considerably.
\\\\

\appendix{\label{ApA}Appendix A}

$$
\eta=\Bigg[4 \Bigg( 2\left(Y +2\dot{H}\right) \left( -6H^{2}+Y
\right) ^{2} \left( 6H^{2}+4\dot{H} +Y  \right) \dddot{H} -8\left(
\frac{3}{2} H^{2}+Y +\frac{5}{2}\dot{H} \right) \left( -6H^{2}+Y
\right) ^{2} \ddot{H}^{2}$$$$-72 \Bigg( 4H\dot{H}^{3}+ \left( 4
H^{3}+\frac{14}{3}Y H -\frac{1}{9}{\frac {\rm d}{{\rm d}t} }Y
\right) \dot{H}^{2}+ \Bigg( \Big( \frac{1}{18}Y- H^{2} \Big) \dot{Y}
+6H^{5}+6Y H^{3}+\frac{7}{6} Y^{2}H \Bigg) \dot{H} $$$$+ \Big(
-\frac{1}{2}H^{4}-\frac{1}{3}H^{2}Y +\frac{1}{24}Y ^{2} \Big)
\dot{Y} +Y H^{3} \left( 6H^{2}+Y \right) \Bigg) \left( -6H^{2}+Y
\right) \ddot{H} + \left( Y +2\dot{H}\right)^{2} \left( Y-6 H^{2}
\right)$$$$ \left( 6H^{2}+4\dot{H} +Y \right) \ddot{Y} + \left( 576
H^{2}-192Y \right) \dot{H}^{5}+ \Bigg( 96 \dot{Y} H-144 \left( 12
H^{2}+Y\right) \left( -4 H^{2}+Y \right) \Bigg) \dot{H}^{4}$$$$+
\Bigg( 5184 H^{6}+4320F H^{4}-1152 Y ^{2} H^{2}-24 Y^{3}+144 \dot{Y}
Y H -4 \dot{Y}^{2} \Bigg) \dot{H} ^{3}+ \Bigg( \left( -12 H^{2}-4Y
\right) \dot{Y}^{2}$$$$+ \Big( 576Y H^{3}-1728 H^{5}+24 Y^{2}H \Big)
\dot{Y} -216Y H^{2} \Big( -12 H^{4}-2 H^{2}Y + Y^{2} \Big) \Bigg)
\dot{H}^{2}-6 \Bigg( \big( 6 H^{4}+\frac{1}{3} Y^{2} \big) \dot{Y}
$$$$ +H \Bigg( 216 H^{6}+36Y H^{4}-30 Y^{2} H^{2}+ Y^{3} \Bigg)
\Bigg) \dot{Y} \dot{H}-\frac{1}{2}\dot{Y}^{2}Y \left( 36
H^{4}-6H^{2}Y + Y^{2} \right) \Bigg) \Bigg]$$$$ \times \Bigg[ \left(
6 H^{2}+4\dot{H} +Y \right) ^{\frac{5}{2}} \left( -6 H^{2}+Y\right)
^{\frac{3}{2}} \left(Y +2\dot{H}\right)^{2}\Bigg]^{-1} \,,
$$

where,
$$
Y=\sum_{i=1}^{3}(\dot{\alpha}^{i})^{2}
$$

\appendix{\label{ApB}Appendix B}

$$
\zeta=\Bigg[\frac{144}{\kappa^{2}} \bigg((2 Y-12 H^{2}) \ddot{H}-24
H\dot{H}^{2}+(2 \dot{Y}-72 H^{3}) \dot{H}+Y \dot{Y}\bigg)
\bigg(-\frac{1}{36}\Big((2 \dot{H}
+Y)^{2}(-6 H^{2}+Y)^{3}$$$$(6 H^{2}+4 \dot{H}+Y)^{2}
(\ddddot{H})\Big)+\Big(\frac{1}{2}\big(\frac{2 H^{2}}{3}+Y+\frac{20
	\dot{H}}{9}\big) \big(Y-6 H^{2} \big)\ddot{H}+\frac{8 H
	\dot{H}^{3}}{3}+\big(\frac{14 Y H}{3}-4 H^{3}\big)
\dot{H}^{2}$$$$+\big(\big(\frac{Y}{3}-2 H^{2}\big) \dot{Y}+12
H^{5}+\frac{5 HY^{2}}{3}\big) \dot{H}+\big(\frac{5
	Y^{2}}{36}-H^{4}-\frac{2 H ^{2} Y}{3}\big) \dot{Y}+H^{3} Y \big(6
H^{2}+Y\big)\Big) \big(2 \dot{H}+Y\big)
$$$$\big(6 H^{2}+4 \dot{H}+Y\big) \big(Y-6 H^{2} \big)^{2} \dddot{H}-\frac{\big(2 \dot{H}+Y \big)^{3} \big(Y-6 H^{2}
	\big)^{2} \big(6 H^{2}+4 \dot{H} +Y\big)^{2}
	\dddot{Y}}{72}-\frac{65}{36}
\Big(\frac{328\dot{H}^{2}}{65}$$$$+\big(\frac{216
	H^{2}}{65}+\frac{292 Y}{65}\big) \dot{H}+\frac{36
	H^{4}}{65}+\frac{96 H^{2} Y}{65}+Y^{2}\Big) \Big(Y-6 H^{2}\Big)^{3}
\ddot{H}^{3}+\Big(Y-6 H^{2}\Big)^{2} \Big(-24 H
\dot{H}^{4}$$$$+\big(176 H ^{3}-\frac{260 Y H
}{3}-\frac{14 \dot{Y}}{9}\big)\dot{H}^{3}+\big(\big(\frac{122 H^{2}}{3}-\frac{82
	Y}{9}\big) \dot{Y}-204 H^{5}+84 H^{3} Y-\frac{175 H Y^{2}}{3}\Big)
\dot{H}^{2}$$$$+\Big(\big(31 H^{4}+\frac{91 H ^{2} Y}{3}-\frac{85
	Y^{2}}{12}\big) \dot{Y}-108 H^{7}-288 Y H^{5}-11 H^{3}
Y^{2}-\frac{26 H Y^{3}}{3}\Big) \dot{H}+\Big(6 H ^{6}+\frac{25 Y H
	^{4}}{2}$$$$+\frac{11 H^{2} Y^{2}}{2}-\frac{107 Y^{3}}{72}\Big)
\dot{Y}-72 Y H^{7}-84 Y^{2} H^{5}-6 Y^{3} H^{3}+H Y^{4}\Big)
\ddot{H}^{2}+9 \Big(\frac{1}{54}\big(6 H ^{2}+4 \dot{H}+Y\big)$$$$
\big(2 \dot{H}+Y\big)^{2} \Big(-6H^{2}+Y\Big) \Big(-2
H^{2}+Y+\frac{4 \dot{H}}{3}\Big) \ddot{Y}+\Big(32 H^{2}+\frac{64 Y
}{27}\Big) \dot{H}^{6}+\Big(\frac{608 H^{4}}{3}+\frac{64 Y
	^{2}}{9}+$$$$\frac{176 H^{2} Y }{3}-\frac{16 H \dot{Y}}{27}\Big)
\dot{H}^{5}+\Big(\frac{164 Y^{3}}{27}-\frac{22
	\dot{Y}^{2}}{81}-\frac{112 (\frac{6 H^{2}}{7}+Y) H \dot{Y}}{27}-240
H ^{6}+496 Y H^{4}+\frac{380 H^{2} Y ^{2}}{9}\Big)
\dot{H}^{4}$$$$+\Big(\big(-\frac{2 H^{2}}{9}-\frac{41 Y}{81}\big)
\dot{Y}^{2}+\big(\frac{248 H^{3} Y}{9}-\frac{242 H Y
	^{2}}{27}-\frac{392 H^{5}}{3}\big) \dot{Y}+192 H ^{8}+2 Y^{4}-64 Y
H^{6}$$$$+\frac{1040 H^{4} Y  ^{2}}{3}+\frac{164 H^{2} Y
	^{3}}{9}\Big) \dot{H}^{3}+\Big(\Big(2 H^{2} Y -7 H^{4}-\frac{59 Y
	^{2}}{108}\Big) \dot{Y}^{2}+\Big(\frac{166 H^{3} Y^{2}}{9}-\frac{316
	Y H^{5}}{3}-\frac{17 H Y^{3}}{3}$$$$-24 H^{7}\Big) \dot{Y}+\frac{2
	Y^{5}}{9}+336 Y H^{8}+52 Y^{2} H^{6}+84 Y^{3} H^{4}+\frac{53 H^{2} Y
	^{4}}{9}\Big) \dot{H}^{2}+\Big(\big(\frac{31 H^{2}
	Y^{2}}{18}-\frac{13 Y H^{4}}{3}-\frac{16 H^{6}}{3}$$$$-\frac{5
	Y^{3}}{18}\big) \dot{Y}^{2}+\Big(20 Y H ^{7}-\frac{100 Y^{2} H
	^{5}}{3}+\frac{5 Y^{3} H^{3}}{9}-\frac{59 H Y ^{4}}{54}+24
H^{9}\Big) \dot{Y}+\Big(6 H^{2}+Y \Big) Y^{2} \Big(20
H^{4}$$$$-\frac{4 H^{2} Y}{3}+Y^{2}\Big) H^{2}\Big)
\dot{H}-\frac{8}{9} \Big(\big(\frac{9 H^{8}}{8}+\frac{9 Y
	H^{6}}{4}+\frac{21 H^{4} Y ^{2}}{32}-\frac{19 H^{2}
	Y^{3}}{48}+\frac{Y^{4}}{18}\big)\dot{Y}-27 Y H^{9}-9 Y^{2}
H^{7}+\frac{21 Y^{3} H^{5}}{4}$$$$+Y ^{4} H^{3}\Big) \dot{Y}\Bigg)
\big(-6 H^{2}+Y \big) \ddot{H}+\frac{1}{6}\bigg(\big(6 H^{2}+4
\dot{H} +Y\big) \big(2 \dot{H} +Y\big)^{2} \big(-6 H^{2}+Y \big)
\Big(-24 H \dot{H}^{3}+$$$$\big(-84 H^{3}-10 Y H+2 \dot{Y}\big)
\dot{H}^{2}+\big(\big(3 H^{2}+\frac{3Y}{2}\big) \dot{Y}+Y H
\big(-42H^{2}+Y\big)\big) \dot{H}+\frac{5}{12}\dot{Y} \big(\frac{72
	H^{4}}{5}$$$$-\frac{6 H^{2} Y}{5}+Y^{2}\big)\Big)
\ddot{Y}\bigg)+\Big(384 H^{3}+768 Y H\Big) \dot{H}^{8}+\Big(4800
H^{3} Y-\big(864 H^{2}+64Y\big) \dot{Y}+1280 H Y^{2}\Big)$$$$
\dot{H}^{7}+\Big(104 H \dot{Y}^{2}-\big(3552 H ^{4}+1760 H^{2} Y
+128 Y^{2}\big) \dot{Y}-19008 H^{7}+14976 Y H^{5}+5232 H^{3}
Y^{2}$$$$+880 H Y^{3}\Big) \dot{H}^{6}+\Big(-\frac{26
	\dot{Y}^{3}}{9}+\big(304 H^{3}+\frac{628 Y H}{3}\big)
\dot{Y}^{2}-\big(\frac{268 Y ^{3}}{3}+4752 H^{6}+1572 H ^{2}
Y^{2}+5280 Y H^{4}\big) \dot{Y}$$$$+5184 H^{9}-24192 Y H^{7}+19440
Y^{2} H ^{5}+1680 Y^{3} H^{3}+328 H Y^{4}\Big)
\dot{H}^{5}+\Big(\big(-6 H ^{2}-\frac{56 Y}{9}\big)
\dot{Y}^{3}+\Big(36 H ^{5}$$$$+\frac{479 H Y^{2}}{3}+596 H^{3}
Y\Big) \dot{Y}^{2}+\Big(11232 H^{8}-11880 Y H^{6}-2088 H ^{4}
Y^{2}-778 H ^{2} Y^{3}-26 Y^{4}\Big) \dot{Y}+5184 Y H^{9}$$$$-9936
Y^{2} H^{7}+8208 Y^{3} H^{5}+36 Y^{4} H^{3}+68 Y ^{5} H\Big)
\dot{H}^{4}+\Big(\Big(H^{4}-\frac{187 Y^{2}}{36}-\frac{37 H ^{2}
	Y}{3}\Big)\dot{Y}^{3}+\Big(756 H^{7}-324 Y H^{5}$$$$+501 H^{3}
Y^{2}+52 H Y    ^{3}\Big)\dot{Y}^{2}+\Big(7776 H ^{10}-\frac{8
	Y^{5}}{3}-211 H^{2} Y^{4}+24 Y^{3} H^{4}+6480 Y H^{8}-7308 Y ^{2}
H^{6}\Big) \dot{Y}$$$$+6 H Y^{2}\big(216 H^{8}-216 Y H^{6}+186 H^{4}
Y^{2}-6 H^{2} Y^{3}+Y ^{4}\big)\Big) \dot{H}^{3}-\frac{1}{2}\Big(49
\Big(\big(-\frac{12 H^{6}}{7}+\frac{39 Y H  ^{4}}{49}+\frac{12 H^{2}
	Y ^{2}}{49}$$$$+\frac{163 Y^{3}}{1764}\big) \dot{Y}^{2}-\frac{10
	\Big(\frac{756H^{8}}{5}+\frac{252 Y H^{6}}{5}-45 H^{4}
	Y^{2}+\frac{192 H^{2} Y^{3}}{5}+Y^{4}\Big) H \dot{Y}}{49}+H^{2} Y
\Big(-\frac{7776 H^{8}}{49}-\frac{864 Y H^{6}}{49}$$$$+\frac{2556
	H^{4} Y ^{2}}{49}-\frac{192 H^{2} Y^{3}}{49}+Y^{4}\Big)\Big)
\dot{Y}\dot{H}^{2}\Big)-\frac{1}{12}\Big(5 \big(\big(\frac{4
	Y^{4}}{3}-\frac{324 H ^{8}}{5}-\frac{288 Y H ^{6}}{5}+\frac{189
	H^{4} Y^{2}}{5}$$$$+\frac{3 H^{2} Y^{3}}{5}\big) \dot{Y}+H
\Big(-\frac{3888 H^{10}}{5}-\frac{1296 Y H^{8}}{5}-\frac{324 Y^{2} H
	^{6}}{5}+\frac{504 Y^{3} H^{4}}{5}-66 H^{2} Y^{4}+Y^{5}\Big)\big)
\dot{Y}^{2} \dot{H} \Big)\bigg)$$$$-\frac{Y \Big(-216 H^{8}-24 Y
	H^{6}+46 H ^{4} Y^{2}-\frac{10 H^{2} Y ^{3}}{3}+Y^{4}\Big)
	\dot{Y}^{3}}{16}\Big]\Big[\Big(2 \dot{H} +Y\Big)^{4} \Big(6 H ^{2}+4
\dot{H}+Y \Big)^{5}$$$$ \Big(-6 H^{2}+Y \Big)^{3}\Big]^{-1}
$$

where,
$$
Y=\sum_{i=1}^{3}(\dot{\alpha}^{i})^{2}.
$$

\appendix{\label{ApC}Appendix C}

$$
\eta=-\Bigg[5184 \Bigg(\frac{b^{\frac{10}{\beta}} \beta^{4}
	\Big(\beta -1\Big)^{3} c^{2} \Big(\beta^{2}-2 \beta
	+\frac{4}{3}\Big) {\mathrm e}^{24 N} N^{\frac{2+5
			\beta}{\beta}}}{18}+\frac{\Big(\beta -\frac{2}{3}\Big) \beta^{2}
	\Big(\beta -1\Big)^{3} c^{4} {\mathrm e}^{18 N} \Big(\beta -2\Big)
	b^{\frac{8}{\beta}} N^{\frac{4+3 \beta}{\beta}}}{216}$$$$+\frac{35
	\beta^{3} \Big(\beta -1\Big)^{2} c^{4} \Big(\beta^{2}-\frac{94}{35}
	\beta +\frac{68}{35}\Big) {\mathrm e}^{18 N} b^{\frac{8}{\beta}}
	N^{\frac{4+4 \beta}{\beta}}}{216}+\frac{11 b^{\frac{10}{\beta}}
	\beta^{5} \Big(\beta -1\Big)^{3} \Big(\beta -\frac{2}{11}\Big) c^{2}
	{\mathrm e}^{24 N} N^{\frac{2+6 \beta}{\beta}}}{18}$$$$+\frac{2
	\Big(\beta^{2}+\frac{1}{2} \beta -2\Big) b^{\frac{10}{\beta}}
	\beta^{6} \Big(\beta -1\Big) c^{2} {\mathrm e}^{24 N} N^{\frac{2+7
			\beta}{\beta}}}{3}+\frac{23 \beta^{4} \Big(\beta -1\Big) c^{4}
	\Big(\beta^{2}-\frac{53}{23} \beta +\frac{31}{23}\Big) {\mathrm
		e}^{18 N} b^{\frac{8}{\beta}} N^{\frac{4+5
			\beta}{\beta}}}{18}$$$$+\frac{b^{\frac{6}{\beta}} c^{6} \beta
	\,{\mathrm e}^{12 N} \Big(\beta -2\Big) \Big(\beta -1\Big)^{2}
	N^{\frac{6+2 \beta}{\beta}}}{216}+\frac{5 b^{\frac{6}{\beta}}
	\beta^{2} \Big(\beta -1\Big) c^{6} {\mathrm e}^{12 N}
	\Big(\beta^{2}-2 \beta +\frac{6}{5}\Big) N^{\frac{3 \beta
			+6}{\beta}}}{108}-\frac{5}{2} b^{\frac{10}{\beta}} \beta^{7}
\Big(\beta -1\Big)$$$$ c^{2} \Big(\beta -\frac{8}{5}\Big) {\mathrm
	e}^{24 N} N^{\frac{2+8 \beta}{\beta}}+\frac{41 \beta^{5} \Big(\beta
	-1\Big) \Big(\beta -\frac{38}{41}\Big) c^{4} {\mathrm e}^{18 N}
	b^{\frac{8}{\beta}} N^{\frac{4+6 \beta}{\beta}}}{12}+\frac{23
	b^{\frac{6}{\beta}} \beta^{3} \Big(\beta -1\Big) c^{6} {\mathrm
		e}^{12 N} \Big(\beta -\frac{36}{23}\Big) N^{\frac{6+4
			\beta}{\beta}}}{72}$$$$+\frac{b^{\frac{4}{\beta}} \beta  \,c^{8}
	{\mathrm e}^{6 N} \Big(6+\beta \Big) \Big(\beta -1\Big) N^{\frac{8+2
			\beta}{\beta}}}{432}-3 b^{\frac{10}{\beta}} c^{2} {\mathrm e}^{24 N}
\beta^{8} \Big(\beta -1\Big) N^{\frac{2+9 \beta}{\beta}}+2
b^{\frac{8}{\beta}} c^{4} \beta^{6} {\mathrm e}^{18 N} \Big(\beta
-1\Big) N^{\frac{4+7 \beta}{\beta}}$$$$+b^{\frac{6}{\beta}} c^{6}
\beta^{4} {\mathrm e}^{12 N} \Big(\beta -1\Big) N^{\frac{6+5
		\beta}{\beta}}-\frac{b^{\frac{4}{\beta}} \beta^{2} c^{8} {\mathrm
		e}^{6 N} \Big(\beta -1\Big) N^{\frac{8+3
			\beta}{\beta}}}{72}+\frac{b^{\frac{2}{\beta}} N^{\frac{2 \beta
			+10}{\beta}} \beta  \,c^{10}}{144}+\frac{3 b^{\frac{6}{\beta}}
	N^{\frac{6+6 \beta}{\beta}} c^{6} \beta^{5} {\mathrm e}^{12
		N}}{4}$$$$-\frac{b^{\frac{4}{\beta}} N^{\frac{4 \beta +8}{\beta}}
	\beta^{3} c^{8} {\mathrm e}^{6 N}}{24}+\Big(\beta -1\Big)
\bigg(-\frac{b^{\frac{4}{\beta}} c^{8} {\mathrm e}^{6 N} \Big(\beta
	-2\Big) \Big(\beta -3\Big)
	N^{\frac{8+\beta}{\beta}}}{1296}+\frac{b^{\frac{2}{\beta}}
	N^{\frac{10+\beta}{\beta}} c^{10}}{432}+\beta^{7} \Big(\beta -1\Big)
N^{7} {\mathrm e}^{30 N} b^{\frac{12}{\beta}}$$$$
\Big(\frac{2}{9}+\Big(N^{2}+\frac{1}{2} N +\frac{1}{6}\Big)
\beta^{2}+\Big(-N -\frac{7}{18}\Big) \beta \Big)\bigg)\Bigg) \beta
\,{\mathrm e}^{6 N}\Bigg]\Bigg[\Big(6 \beta^{2} N^{2}
b^{\frac{2}{\beta}} {\mathrm e}^{6 N}-c^{2}
N^{\frac{2}{\beta}}\Big)^{\frac{3}{2}} \Big(6 \beta  N \,{\mathrm
	e}^{6 N} \Big((N +\frac{2}{3}) \beta-\frac{2}{3} \Big)$$$$
b^{\frac{2}{\beta}}+c^{2} N^{\frac{2}{\beta}}\Big)^{\frac{5}{2}}
\Big(2 N \beta  \,{\mathrm e}^{6 N} \Big(\beta -1\Big)
b^{\frac{2}{\beta}}+c^{2} N^{\frac{2}{\beta}}\Big)^{2}\Bigg]^{-1}
$$

\appendix{\label{ApD}Appendix D}
$$
\zeta=\Bigg[{\mathrm e}^{12 N} \Big(-10 b^{\frac{6}{\beta}} c^{18}
{\mathrm e}^{6 N} \beta  (\beta -2) (\beta -3) N^{\frac{18+3
		\beta}{\beta}}+\frac{b^{\frac{8}{\beta}} c^{16} {\mathrm e}^{12 N}
	(\beta -1) (\beta -3) (\beta -4) (\beta -2)^{2} N^{\frac{16+2
			\beta}{\beta}}}{9}$$$$+5544 c^{16} b^{\frac{8}{\beta}} \Big(\beta
-\frac{113}{77}\Big) {\mathrm e}^{12 N} \beta^{4} N^{\frac{16+6
		\beta}{\beta}}-\frac{2 b^{\frac{6}{\beta}} c^{18} {\mathrm e}^{6 N}
	\Big(\beta -2\Big) \Big(\beta^{2}-5 \beta +7\Big) N^{\frac{18+2
			\beta}{\beta}}}{3}+203148 c^{12} {\mathrm e}^{24 N}
\Big(\beta^{3}$$$$-\frac{16583}{5643} \beta^{2}+\frac{16540}{5643}
\beta -\frac{5576}{5643}\Big) b^{\frac{12}{\beta}} \beta^{6}
N^{\frac{8 \beta +12}{\beta}}-200 b^{\frac{16}{\beta}} c^{8}
\Big(\beta -2\Big) \Big(\beta^{3}-\frac{14}{5}
\beta^{2}+\frac{108}{25} \beta -\frac{56}{15}\Big) {\mathrm e}^{36
	N} \beta^{5} \Big(\beta -1\Big)^{4}$$$$ N^{\frac{8+7
		\beta}{\beta}}+164304 b^{\frac{16}{\beta}} c^{8} {\mathrm e}^{36 N}
\beta^{7} \Big(\beta^{4}-\frac{679}{163} \beta^{3}+\frac{7428}{1141}
\beta^{2}-\frac{15544}{3423} \beta +\frac{1384}{1141}\Big)
\Big(\beta -1\Big)^{2} N^{\frac{8+9 \beta}{\beta}}-\frac{1088}{3}
b^{\frac{18}{\beta}} c^{6} \Big(\beta -2\Big)$$$$ \beta^{6} {\mathrm
	e}^{42 N} \Big(\beta -1\Big)^{5} \Big(\beta^{3}-\frac{70}{17}
\beta^{2}+\frac{108}{17} \beta -\frac{64}{17}\Big) N^{\frac{6+8
		\beta}{\beta}}-5396 b^{\frac{16}{\beta}}
\Big(\beta^{3}-\frac{2922}{1349} \beta^{2}+\frac{388}{1349} \beta
+\frac{72}{71}\Big) c^{8} \Big(\beta -2\Big) {\mathrm e}^{36 N}
\beta^{6}$$$$ \Big(\beta -1\Big)^{3} N^{\frac{8+8
		\beta}{\beta}}-7264 b^{\frac{18}{\beta}} c^{6}
\Big(\beta^{3}-\frac{722}{227} \beta^{2}+\frac{616}{227} \beta
-\frac{64}{227}\Big) \Big(\beta -2\Big) \beta^{7} {\mathrm e}^{42 N}
\Big(\beta -1\Big)^{4} N^{\frac{6+9 \beta}{\beta}}+8
b^{\frac{16}{\beta}} c^{8} \Big(\beta -2\Big)^{2}$$$$
\Big(\beta^{2}-\frac{20}{9} \beta +\frac{20}{9}\Big) {\mathrm e}^{36
	N} \beta^{4} (\beta -1)^{5} N^{\frac{8+6
		\beta}{\beta}}+N^{\frac{20+2 \beta}{\beta}} c^{20} (\beta -2)
b^{\frac{4}{\beta}}-559872
\Big(\frac{64}{81}+\Big(N^{3}+\frac{53}{18} N^{2}+\frac{13}{9} N
+\frac{7}{27}\Big) \beta^{4}$$$$+\Big(-\frac{8}{3}
N^{3}-\frac{22}{3} N^{2}-\frac{55}{9} N -\frac{118}{81}\Big)
\beta^{3}+\Big(\frac{16}{3} N^{2}+\frac{74}{9} N +\frac{79}{27}\Big)
\beta^{2}+\Big(-\frac{32 N}{9}-\frac{68}{27}\Big) \beta \Big)
b^{\frac{24}{\beta}} N^{14} \beta^{13} {\mathrm e}^{60 N} (\beta
-1)^{3}$$$$ \Big(-\frac{2}{3}+\Big(N +\frac{1}{2}\Big) \beta
\Big)\Big) \Big(\beta -1\Big)+960 \Big(\beta -4\Big) c^{4}
\Big(\beta -2\Big)^{2} \Big(\beta -\frac{8}{5}\Big) \beta^{8}
{\mathrm e}^{48 N} b^{\frac{20}{\beta}} \Big(\beta -1\Big)^{6}
N^{\frac{4+10 \beta}{\beta}}-\frac{2}{3} b^{\frac{10}{\beta}} c^{14}
{\mathrm e}^{18 N} \beta $$$$ \Big(2+\beta \Big) \Big(\beta -3\Big)
\Big(\beta -2\Big)^{2} \Big(\beta -1\Big)^{3} N^{\frac{14+3
		\beta}{\beta}}+7838208 b^{\frac{18}{\beta}} c^{6} {\mathrm e}^{42 N}
\beta^{14} \Big(\beta -1\Big)^{2} N^{\frac{6+16
		\beta}{\beta}}-1119744 b^{\frac{22}{\beta}} c^{2} {\mathrm e}^{54 N}
\beta^{17} $$$$\Big(\beta -1\Big)^{3} N^{\frac{2+18
		\beta}{\beta}}+1679616 b^{\frac{20}{\beta}} c^{4} \beta^{16}
{\mathrm e}^{48 N} \Big(\beta -1\Big)^{2} N^{\frac{4+18
		\beta}{\beta}}+21928320 b^{\frac{18}{\beta}} c^{6} \beta^{12}
\Big(\beta^{2}-\frac{1502}{705} \beta +\frac{812}{705}\Big) {\mathrm
	e}^{42 N}$$$$ \Big(\beta -1\Big)^{2} N^{\frac{6+14
		\beta}{\beta}}+31850496 \Big(\beta^{3}-\frac{799}{256}
\beta^{2}+\frac{1693}{512} \beta -\frac{153}{128}\Big) c^{4}
\beta^{13} {\mathrm e}^{48 N} b^{\frac{20}{\beta}} \Big(\beta
-1\Big)^{2} N^{\frac{4+15 \beta}{\beta}}+13115520
b^{\frac{18}{\beta}} c^{6} $$$$\Big(\beta^{3}-\frac{3934}{1265}
\beta^{2}+\frac{4083}{1265} \beta -\frac{284}{253}\Big) \beta^{11}
{\mathrm e}^{42 N} \Big(\beta -1\Big)^{2} N^{\frac{6+13
		\beta}{\beta}}+3234816 b^{\frac{16}{\beta}} c^{8} {\mathrm e}^{36 N}
\beta^{9} \Big(\beta -1\Big)^{2} \Big(\beta^{3}-\frac{1325}{468}
\beta^{2}$$$$+\frac{4837}{1872} \beta -\frac{173}{234}\Big)
N^{\frac{8+11 \beta}{\beta}}+653472 \,{\mathrm e}^{30 N}
b^{\frac{14}{\beta}} c^{10} \Big(\beta^{3}-\frac{6812}{2269}
\beta^{2}+\frac{6899}{2269} \beta -\frac{2352}{2269}\Big) \beta^{7}
\Big(\beta -1\Big)^{2} N^{\frac{10+9 \beta}{\beta}}+30000$$$$ c^{12}
{\mathrm e}^{24 N} b^{\frac{12}{\beta}}
\Big(\beta^{3}-\frac{399}{125} \beta^{2}+\frac{2153}{625} \beta
-\frac{162}{125}\Big) \beta^{5} \Big(\beta -1\Big)^{2} N^{\frac{12+7
		\beta}{\beta}}+17879616 c^{4} \Big(\beta^{4}-\frac{14470}{3449}
\beta^{3}+\frac{22895}{3449} \beta^{2}$$$$-\frac{16248}{3449} \beta
+\frac{4380}{3449}\Big) \beta^{12} {\mathrm e}^{48 N}
b^{\frac{20}{\beta}} \Big(\beta -1\Big)^{2} N^{\frac{4+14
		\beta}{\beta}}+1229760 b^{\frac{16}{\beta}} c^{8}
\Big(\beta^{4}-\frac{32769}{8540} \beta^{3}+\frac{9259}{1708}
\beta^{2}-\frac{7094}{2135} \beta +\frac{1576}{2135}\Big)
$$$${\mathrm e}^{36 N} \beta^{8} \Big(\beta -1\Big)^{2} N^{\frac{8+10
		\beta}{\beta}}+1628784 c^{4} \beta^{10} {\mathrm e}^{48 N}
b^{\frac{20}{\beta}} \Big(\beta^{4}-\frac{48536}{11311}
\beta^{3}+\frac{78064}{11311} \beta^{2}-\frac{55120}{11311} \beta
+\frac{14144}{11311}\Big) \Big(\beta -1\Big)^{4}$$$$ N^{\frac{4+12
		\beta}{\beta}}+919296 b^{\frac{22}{\beta}}
\Big(\beta^{4}-\frac{477}{266} \beta^{3}-\frac{103}{133}
\beta^{2}+\frac{362}{133} \beta -\frac{8}{7}\Big) c^{2} {\mathrm
	e}^{54 N} \beta^{12} \Big(\beta -1\Big)^{4} N^{\frac{2+14
		\beta}{\beta}}+198528 b^{\frac{18}{\beta}} c^{6}
\Big(\beta^{4}$$$$-\frac{4115}{1034} \beta^{3}+\frac{2947}{517}
\beta^{2}-\frac{1761}{517} \beta +\frac{360}{517}\Big) \beta^{8}
{\mathrm e}^{42 N} \Big(\beta -1\Big)^{4} N^{\frac{6+10
		\beta}{\beta}}+1752192 b^{\frac{22}{\beta}} c^{2}
\Big(\beta^{4}-\frac{181}{169} \beta^{3}-\frac{426}{169}
\beta^{2}+\frac{648}{169} \beta$$$$ -\frac{200}{169}\Big) {\mathrm
	e}^{54 N} \beta^{13} \Big(\beta -1\Big)^{3} N^{\frac{2+15
		\beta}{\beta}}+1664064 \,{\mathrm e}^{30 N} b^{\frac{14}{\beta}}
\Big(\beta -\frac{105}{107}\Big) c^{10} \beta^{10} \Big(\beta
-1\Big) N^{\frac{10+12 \beta}{\beta}}+364176 c^{12} {\mathrm e}^{24
	N} b^{\frac{12}{\beta}}$$$$ \beta^{8} \Big(\beta
-\frac{301}{281}\Big) (\beta -1) N^{\frac{12+10 \beta}{\beta}}+84024
c^{14} \Big(\beta -\frac{314}{389}\Big) {\mathrm e}^{18 N} \beta^{6}
b^{\frac{10}{\beta}} (\beta -1) N^{\frac{14+8 \beta}{\beta}}-18
b^{\frac{6}{\beta}} c^{18} {\mathrm e}^{6 N} \beta^{2} (\beta -1)
(\beta -8)$$$$ N^{\frac{18+4 \beta}{\beta}}+91272 \,{\mathrm e}^{30
	N} b^{\frac{14}{\beta}} c^{10} \beta^{6} \Big(\beta -1\Big)^{2}
\Big(\beta^{4}-\frac{16639}{3803} \beta^{3}+\frac{27936}{3803}
\beta^{2}-\frac{21372}{3803} \beta +\frac{6296}{3803}\Big)
N^{\frac{10+8 \beta}{\beta}}+2083968 b^{\frac{22}{\beta}} c^{2}$$$$
{\mathrm e}^{54 N} \beta^{14} \Big(\beta^{3}+\frac{16}{67}
\beta^{2}-\frac{176}{67} \beta +\frac{64}{67}\Big) \Big(\beta
-1\Big)^{3} N^{\frac{2+16 \beta}{\beta}}+1615968
b^{\frac{18}{\beta}} c^{6} \beta^{9} {\mathrm e}^{42 N}
\Big(\beta^{4}-\frac{21671}{5611} \beta^{3}+\frac{30606}{5611}
\beta^{2}$$$$-\frac{18436}{5611} \beta +\frac{3856}{5611}\Big)
\Big(\beta -1\Big)^{3} N^{\frac{6+11 \beta}{\beta}}+6426432 c^{4}
\beta^{11} {\mathrm e}^{48 N} \Big(\beta^{4}-\frac{15779}{3719}
\beta^{3}+\frac{25320}{3719} \beta^{2}-\frac{18120}{3719} \beta
+\frac{4848}{3719}\Big) b^{\frac{20}{\beta}}$$$$ (\beta -1)^{3}
N^{\frac{4+13 \beta}{\beta}}+4898880 \Big(\beta -\frac{16}{15}\Big)
b^{\frac{16}{\beta}} c^{8} {\mathrm e}^{36 N} \beta^{12} (\beta -1)
N^{\frac{8+14 \beta}{\beta}}+208320 c^{4} \beta^{9} {\mathrm e}^{48
	N} b^{\frac{20}{\beta}} (\beta -1)^{5} \Big(\beta^{4}-\frac{32}{7}
\beta^{3}$$$$+\frac{8444}{1085} \beta^{2}-\frac{6256}{1085} \beta
+\frac{1664}{1085}\Big) N^{\frac{4+11 \beta}{\beta}}+343872
b^{\frac{22}{\beta}} c^{2} \Big(\beta^{4}-\frac{1856}{597}
\beta^{3}+\frac{1600}{597} \beta^{2}+\frac{32}{199} \beta
-\frac{448}{597}\Big) {\mathrm e}^{54 N} \beta^{11} (\beta
-1)^{5}$$$$ N^{\frac{2+13 \beta}{\beta}}+5403456
b^{\frac{18}{\beta}} c^{6} \Big(\beta^{4}-\frac{12466}{3127}
\beta^{3}+\frac{18567}{3127} \beta^{2}-\frac{12146}{3127} \beta
+\frac{2912}{3127}\Big) \beta^{10} {\mathrm e}^{42 N} \Big(\beta
-1\Big)^{2} N^{\frac{6+12 \beta}{\beta}}+8040 c^{14}$$$$ {\mathrm
	e}^{18 N} \beta^{4} b^{\frac{10}{\beta}}
\Big(\beta^{3}-\frac{1106}{335} \beta^{2}+\frac{1199}{335} \beta
-\frac{434}{335}\Big) \Big(\beta -1\Big) N^{\frac{14+6
		\beta}{\beta}}+30839616 c^{4} \Big(\beta^{2}-\frac{1408}{661} \beta
+\frac{768}{661}\Big) \beta^{14} {\mathrm e}^{48 N}
b^{\frac{20}{\beta}} $$$$\Big(\beta -1\Big)^{2} N^{\frac{4+16
		\beta}{\beta}}+5444496 b^{\frac{16}{\beta}} c^{8} {\mathrm e}^{36 N}
\beta^{10} \Big(\beta^{3}-\frac{12141}{4201}
\beta^{2}+\frac{11684}{4201} \beta -\frac{3756}{4201}\Big)
\Big(\beta -1\Big) N^{\frac{8+12 \beta}{\beta}}+1977696 \,{\mathrm
	e}^{30 N} $$$$b^{\frac{14}{\beta}} c^{10}
\Big(\beta^{2}-\frac{1489}{763} \beta +\frac{732}{763}\Big)
\beta^{9} (\beta -1) N^{\frac{10+11 \beta}{\beta}}+13996800 c^{4}
\beta^{15} {\mathrm e}^{48 N} b^{\frac{20}{\beta}} \Big(\beta
-\frac{28}{25}\Big) (\beta -1)^{2} N^{\frac{4+17
		\beta}{\beta}}+6578496 $$$$b^{\frac{16}{\beta}} c^{8} {\mathrm
	e}^{36 N} \beta^{11} \Big(\beta^{2}-\frac{865}{423} \beta
+\frac{446}{423}\Big) \Big(\beta -1\Big) N^{\frac{8+13
		\beta}{\beta}}+1212 c^{16} b^{\frac{8}{\beta}} {\mathrm e}^{12 N}
\beta^{3} \Big(\beta^{2}-\frac{261}{101} \beta +\frac{180}{101}\Big)
\Big(\beta -1\Big) N^{\frac{16+5 \beta}{\beta}}$$$$+44784 c^{14}
\Big(\beta^{2}-2 \beta +\frac{313}{311}\Big) {\mathrm e}^{18 N}
\beta^{5} b^{\frac{10}{\beta}} \Big(\beta -1\Big) N^{\frac{14+7
		\beta}{\beta}}+20528640 \Big(\beta -\frac{12}{11}\Big)
b^{\frac{18}{\beta}} c^{6} \beta^{13} {\mathrm e}^{42 N} \Big(\beta
-1\Big)^{2} N^{\frac{6+15 \beta}{\beta}}$$$$+419472 c^{12} {\mathrm
	e}^{24 N} \Big(\beta^{2}-\frac{1893}{971} \beta
+\frac{898}{971}\Big) b^{\frac{12}{\beta}} \beta^{7} \Big(\beta
-1\Big) N^{\frac{12+9 \beta}{\beta}}+1541376 \,{\mathrm e}^{30 N}
b^{\frac{14}{\beta}} c^{10} \Big(\beta^{3}-\frac{2665}{892}
\beta^{2}+\frac{5221}{1784} \beta $$$$ -\frac{209}{223}\Big)
\beta^{8} (\beta -1) N^{\frac{10+10 \beta}{\beta}}+2612736
b^{\frac{22}{\beta}} c^{2} \Big(\beta -\frac{2}{7}\Big) {\mathrm
	e}^{54 N} \beta^{15} (\beta -1)^{3} N^{\frac{2+17 \beta}{\beta}}+18
b^{\frac{4}{\beta}} N^{\frac{20+4 \beta}{\beta}} c^{20}
\beta^{2}-252 b^{\frac{6}{\beta}} c^{18}$$$$ {\mathrm e}^{6 N}
\beta^{3} \Big(\beta -1\Big) N^{\frac{18+5 \beta}{\beta}}+1399680
b^{\frac{16}{\beta}} c^{8} {\mathrm e}^{36 N} \beta^{13} \Big(\beta
-1\Big) N^{\frac{8+15 \beta}{\beta}}+699840 b^{\frac{14}{\beta}}
c^{10} {\mathrm e}^{30 N} \beta^{11} \Big(\beta -1\Big)
N^{\frac{10+13 \beta}{\beta}}$$$$+241056 b^{\frac{12}{\beta}} c^{12}
{\mathrm e}^{24 N} \beta^{9} \Big(\beta -1\Big) N^{\frac{12+11
		\beta}{\beta}}+32400 b^{\frac{10}{\beta}} c^{14} {\mathrm e}^{18 N}
\beta^{7} \Big(\beta -1\Big) N^{\frac{14+9 \beta}{\beta}}+12744
b^{\frac{8}{\beta}} c^{16} {\mathrm e}^{12 N} \beta^{5} \Big(\beta
-1\Big)$$$$ N^{\frac{16+7 \beta}{\beta}}+104976 b^{\frac{12}{\beta}}
N^{\frac{12+12 \beta}{\beta}} c^{12} \beta^{10} {\mathrm e}^{24
	N}+12 b^{\frac{4}{\beta}} c^{20} \beta  \Big(\beta -1\Big)
N^{\frac{20+3 \beta}{\beta}}+8748 b^{\frac{8}{\beta}} N^{\frac{16+8
		\beta}{\beta}} c^{16} \beta^{6} {\mathrm e}^{12 N}-11664
b^{\frac{10}{\beta}}$$$$ N^{\frac{14+10 \beta}{\beta}} c^{14}
\beta^{8} {\mathrm e}^{18 N}-324 b^{\frac{6}{\beta}} N^{\frac{18+6
		\beta}{\beta}} c^{18} \beta^{4} {\mathrm e}^{6 N}-\frac{8}{3} c^{14}
\Big(\beta^{3}-\frac{15}{2} \beta^{2}+\frac{77}{2} \beta -35\Big)
(\beta -2) {\mathrm e}^{18 N} \beta^{2} b^{\frac{10}{\beta}} (\beta
-1)^{2}$$$$ N^{\frac{14+4 \beta}{\beta}}+57600 b^{\frac{22}{\beta}}
c^{2} {\mathrm e}^{54 N} \beta^{10} \Big(\beta -\frac{4}{3}\Big)
\Big(\beta -1\Big)^{6} \Big(\beta^{3}-\frac{68}{25}
\beta^{2}+\frac{44}{25} \beta +\frac{8}{25}\Big) N^{\frac{2+12
		\beta}{\beta}}+\frac{4}{9} b^{\frac{14}{\beta}} c^{10} {\mathrm
	e}^{30 N} \beta^{3} \Big(\beta^{2}$$$$+20 \beta +4\Big) (\beta
-2)^{2} (\beta -1)^{5} N^{\frac{10+5 \beta}{\beta}}+47 c^{16}
b^{\frac{8}{\beta}} \Big(\beta^{2}-\frac{97}{47} \beta
+\frac{62}{47}\Big) (\beta -2) {\mathrm e}^{12 N} \beta^{2} (\beta
-1) N^{\frac{16+4 \beta}{\beta}}+48 c^{14} (\beta -2)$$$$ {\mathrm
	e}^{18 N} \beta^{3} b^{\frac{10}{\beta}} \Big(\beta^{2}+\frac{1}{2}
\beta -\frac{11}{6}\Big) (\beta -1)^{2} N^{\frac{14+5
		\beta}{\beta}}+4 \Big(\beta^{2}-3 \beta -\frac{1}{3}\Big) c^{16}
b^{\frac{8}{\beta}} (\beta -2) {\mathrm e}^{12 N} \beta (\beta
-1)^{2} N^{\frac{16+3 \beta}{\beta}}-\frac{31}{9} c^{12} {\mathrm
	e}^{24 N} $$$$\Big(\beta^{2}-\frac{116}{31} \beta +\frac{4}{31}\Big)
b^{\frac{12}{\beta}} (\beta -2)^{2} \beta^{2} (\beta -1)^{4}
N^{\frac{12+4 \beta}{\beta}}-\frac{464}{3} {\mathrm e}^{30 N}
b^{\frac{14}{\beta}} c^{10} (\beta -2) \beta^{4}
\Big(\beta^{3}-\frac{223}{58} \beta^{2}+\frac{220}{29} \beta
-\frac{168}{29}\Big)$$$$ (\beta -1)^{4} N^{\frac{10+6
		\beta}{\beta}}-\frac{284 c^{12} {\mathrm e}^{24 N}
	b^{\frac{12}{\beta}} (\beta -2) \Big(\beta^{3}-\frac{341}{71}
	\beta^{2}+\frac{694}{71} \beta -\frac{468}{71}\Big) \beta^{3} (\beta
	-1)^{3} N^{\frac{12+5 \beta}{\beta}}}{3}-2696 \,{\mathrm e}^{30 N}
\Big(\beta^{3}$$$$-\frac{683}{337} \beta^{2}+\frac{8}{337} \beta
+\frac{344}{337}\Big) b^{\frac{14}{\beta}} c^{10} \Big(\beta -2\Big)
\beta^{5} \Big(\beta -1\Big)^{3} N^{\frac{10+7 \beta}{\beta}}-1068
c^{12} {\mathrm e}^{24 N} b^{\frac{12}{\beta}} \Big(\beta -2\Big)
\beta^{4} \Big(\beta^{3}-\frac{926}{267} \beta^{2}+\frac{329}{89}
\beta $$$$-\frac{334}{267}\Big) \Big(\beta -1\Big)^{2} N^{\frac{12+6
		\beta}{\beta}}\Big) \beta^{2}\Bigg]\Bigg[373248 \Big(\beta  N
\,{\mathrm e}^{6 N} \Big(\beta -1\Big)
b^{\frac{2}{\beta}}+\frac{c^{2} N^{\frac{2}{\beta}}}{2}\Big)^{4}
\Big(\beta^{2} N^{2} {\mathrm e}^{6 N}
b^{\frac{2}{\beta}}-\frac{c^{2} N^{\frac{2}{\beta}}}{6}\Big)^{3}
\Big({\mathrm e}^{6 N} N \beta $$$$ \Big(-\frac{2}{3}+\Big(N
+\frac{2}{3}\Big) \beta \Big) b^{\frac{2}{\beta}}+\frac{c^{2}
	N^{\frac{2}{\beta}}}{6}\Big)^{5}\Bigg]
$$

\end{document}